\documentclass[11pt, cls, onecolumn]{IEEEtran}

\usepackage{enumitem,xcolor}
\usepackage{subfigure,cite,epsfig,dblfloatfix}
\usepackage[cmex10]{amsmath}
\usepackage{amssymb,mathrsfs,graphicx}
\usepackage{psfrag}

\DeclareFontEncoding{U}{}{}
\DeclareFontFamily{U}{bbold}{}
\DeclareFontShape{U}{bbold}{m}{n}
 {  <5> <6> <7> <8> <9> gen * bbold
   <10> <10.95> bbold10
  <12> <14.4> bbold12
 <17.28> <20.74> <24.88> bbold17
  }{}
\DeclareSymbolFont{bbold}{U}{bbold}{m}{n}
\DeclareSymbolFontAlphabet{\mathbbold}{bbold}

\let\oldbrace\{
\def\{{\oldbrace\kern0.5pt}

\def\tr{\mathop{\rm tr}\nolimits}%
\newtheorem{theorem}{Theorem}
\newtheorem{example}{Example}
\newtheorem{remark}{Remark}
\newtheorem{lemma}{Lemma}
\newtheorem{proposition}{Proposition}
\newtheorem{corollary}{Corollary}

\newcommand{\markov}{\mathrel\multimap\joinrel\mathrel-\mspace{-9mu}\joinrel\mathrel-}

\begin{document}
\title{On Achievability for Downlink Cloud Radio Access Networks with Base Station Cooperation}
\author{Chien-Yi Wang, Mich\`{e}le Wigger, and Abdellatif Zaidi
\thanks{The work of C.-Y. Wang and M. Wigger has been supported by Huawei Technologies France SASU, under grant agreement YB2015120036.}
\thanks{C.-Y. Wang and M. Wigger are with the Communications and Electronics Department, Telecom ParisTech, Universit\'{e} Paris-Saclay, Paris, France. Emails: \{chien-yi.wang, michele.wigger\}@telecom-paristech.fr}
\thanks{A. Zaidi is with the Mathematics and Algorithmic Sciences Lab, Huawei Technologies France, Boulogne-Billancourt, France. Email: abdellatif.zaidi@huawei.com}
}

\maketitle

\begin{abstract}
This work investigates the downlink of a cloud radio access network (C-RAN) in which a central processor communicates with two mobile users through two base stations (BSs). The BSs act as relay nodes and cooperate with each other through error-free rate-limited links. We develop and analyze two coding schemes for this scenario. The first coding scheme is based on Liu--Kang scheme for C-RANs without BS cooperation; and extends it to scenarios allowing conferencing between the BSs. Among few other features, our new coding scheme enables arbitrary correlation among the auxiliary codewords that are recovered by the BSs. It also introduces common codewords to be described to both BSs. For the analysis of this coding scheme, we extend the multivariate covering lemma to non-Cartesian product sets, thereby correcting an erroneous application of this lemma in Liu--Kang's related work. We highlight key aspects of this scheme by studying three important instances of it. The second coding scheme extends the so-called compression scheme that was originally developed for memoryless Gaussian C-RANs without BS cooperation to general discrete memoryless C-RANs with BS cooperation. We show that this scheme subsumes the original compression scheme when applied to memoryless Gaussian C-RAN models. In the analysis of this scheme, we also highlight important connections with the so-called distributed decode--forward scheme, and refine the approximate capacity of a general $N$-BS $L$-user C-RAN model in the memoryless Gaussian case.

\end{abstract}


\section{Introduction} \label{sec:intro}

Cloud radio access networks (C-RANs) are promising candidates for fifth generation (5G) wireless communication networks. In a C-RAN, the base stations (BSs) are connected to a central processor through digital fronthaul links. Comprehensive surveys on C-RANs can be found in \cite{Simeone:16,Peng:16}. The $2$-BS $2$-user case is depicted in Figure~\ref{fig:system}. The two most important coding schemes for downlink C-RANs are  
\begin{itemize}[leftmargin=*]
\item The \emph{data-sharing} scheme: The central processor splits each message into independent submessages and conveys these independent submessages to one or multiple BSs. The BSs map the received submessages into codewords and transmit these codewords over the interference network. The mobile users decode their intended message parts by treating interference as noise. 
If there are $N$ BSs, in general there can be up to $2^N-1$ submessages, each of which is sent to a specific subset of BSs. Two special cases have been considered in the literature: Zakhour and Gesbert~\cite{Zakhour:11} studied the $2$-BS $2$-user case. On the other hand, Dai and Yu~\cite{Dai:14} focused on BS clustering for general C-RANs: The messages are sent as a whole to subsets of BSs and there is no message splitting.

\begin{figure}[t!]
\begin{center}
\includegraphics[scale=0.8, trim={0.5cm 0.5cm 5cm 24cm}, clip]{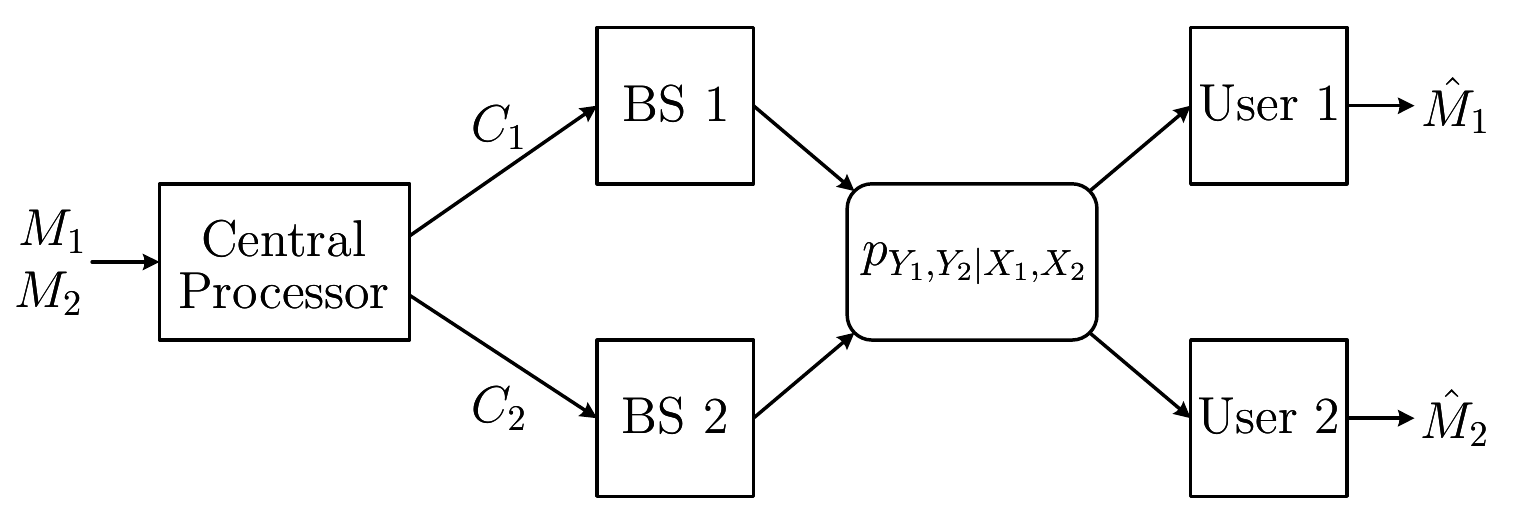}
\end{center}
\vspace{-0.15in}
\caption{Downlink C-RAN with $2$ base stations and $2$ mobile users.}
\label{fig:system}
\vspace{-0.1in}
\end{figure}

\item The \emph{compression} scheme: The central processor first precalculates idealized channel inputs and then sends lossy representations of these idealized inputs over the rate-limited fronthaul links to the BSs. The BSs reconstruct the compressed signals and transmit them over the interference network. 
The first hop, from the central processor to the BSs, is conceptually a lossy source coding problem. The goal is to make the compressed signals correlated in a way that would be useful for the second hop, from the BSs to the mobile users. The compression scheme was first investigated by Park~{\em et al.}~\cite{Park:13} for the memoryless Gaussian case. 
\end{itemize}
A third scheme, the \emph{reverse compute--forward}, was proposed by Hong and Caire~\cite{Hong:13}, which uses nested lattice codes to perform precalculations in a finite field.
The reverse compute--forward scheme can enhance the performance under the condition of weak fronthaul links, but it suffers from non-integer penalty and thus is less competitive than the first two schemes when the fronthaul links are strong.

Recently, for the downlink of C-RANs some advanced coding schemes have been developed based on random coding: Liu and Kang \cite{Liu:14} generalized the data-sharing scheme to a new scheme, which we will refer to as Liu--Kang scheme. In the Liu--Kang scheme, the central processor maps the message pair $(M_1, M_2)$ into ``$2$-dimensional" Marton codewords: codewords $U_1^n, U_2^n$ for message $M_1$ and $V_1^n, V_2^n$ for message $M_2$. The central processor then describes codewords $U_1^n, V_1^n$ to BS~$1$ and codewords $U_2^n, V_2^n$ to BS~$2$, where the descriptions are obtained by enumerating all possible pairs of codewords $(U_1^n, V_1^n)$ and $(U_2^n, V_2^n)$. However, the performance analysis in \cite{Liu:14} is flawed due to an erroneous application of the mutual covering lemma. This leads to a rate region that is not achievable using the described coding scheme, because of some missing rate constraints.
 
On the other hand, it was observed in~\cite{Yu:16} that for the $2$-BS $2$-user case, distributed decode--forward (DDF)~\cite{Lim:15} subsumes the compression scheme. The DDF scheme precodes every codeword involved in the entire communication already at the source (the central processor, in our setup). The codewords carry the information of the messages in an implicit manner. 

In this paper, we study the downlink of a C-RAN with two BSs and two mobile users in which the BSs cooperate over error-free rate-limited links. We develop two coding schemes for this model. The first coding scheme, termed {\em generalized data-sharing (G-DS)}, is based on a variation of the Liu--Kang scheme~\cite{Liu:14}, which is developed for C-RANs without BS cooperation. Our G-DS scheme accounts for the conferencing between the BSs by introducing common codewords ($U_0^n,V_0^n$) intended to be recovered by both of them. The analysis generalizes that of~\cite{Liu:14} and fixes an erroneous step in the achievability proof therein. To this end, in particular we extend the multivariate covering lemma to non-Cartesian product sets. 

The second coding scheme, termed {\em generalized compression (G-Compression)}, is based on the compression scheme developed by Park~{\em et al.}~\cite{Park:13} in the context of Gaussian C-RAN without BS cooperation. Our G-Compression scheme also accounts for conferencing between the BSs and applies to general discrete memoryless channels on the second hop. We analyze this scheme and show that its performance subsumes that of  
the DDF scheme when the latter is applied to the studied downlink C-RAN model. Furthermore, we characterize the capacity region of a general $N$-BS $L$-user C-RAN model under the memoryless Gaussian model to within a better (i.e., smaller) constant gap, independent of power.

The main contributions of this work can be summarized as follows:
\begin{enumerate}[leftmargin=*]
\item We modify the Liu--Kang scheme~\cite{Liu:14} and introduce common codewords to the new G-DS scheme. We use the cooperation links to exchange part of common codewords and to redirect private codewords for asymmetric link or channel conditions. The new G-DS scheme subsumes the data-sharing scheme proposed in~\cite{Zakhour:11}. To highlight distinct features of the code components, we consider three representative simplifications. 
\item We introduce a cloud center and incorporate BS cooperation to the compression scheme in~\cite{Park:13} and derive the corresponding achievable rate region for general discrete memoryless channels on the second hop. The new G-Compression scheme subsumes the scheme proposed in~\cite{Park:13} when adapted to the memoryless Gaussian case. 
\item We simplify the achievable rate region of the DDF scheme for downlink C-RAN with BS cooperation. Under the memoryless Gaussian model, we characterize the capacity region of a downlink $N$-BS $L$-user C-RAN with BS cooperation to within a gap of $\frac{L}{2}+\frac{\min\{N,L\log N\}}{2}$ bits per dimension, which improves the previous result $\frac{L+N}{2}$.
\item We show that under the memoryless Gaussian model, the G-DS scheme outperforms the G-Compression scheme in the low-power regime and when the channel gain matrix is ill-conditioned. Furthermore, compared to the G-Compression scheme, the G-DS scheme benefits more from BS cooperation. 
\end{enumerate}

The paper is organized as follows. In Section~\ref{sec:prob_state}, we provide the problem formulation for the $2$-BS $2$-user case. Section~\ref{sec:description} is devoted to the G-DS scheme, in which we describe the detailed coding scheme and consider three representative special cases and two examples with simpler network topologies. Section~\ref{sec:compress} is devoted to the G-Compression scheme. In this section, we describe the G-Compression scheme with a cloud center and then conduct a performance analysis on the DDF scheme. Finally, in Section~\ref{sec:compare_evaluate} we compare the G-DS scheme and the G-Compression scheme through examples and evaluation for the memoryless Gaussian model. The lengthy proofs are deferred to appendices.

\subsection{Notations} 
Random variables and their realizations are represented by uppercase letters (e.g., $X$) and lowercase letters (e.g., $x$), respectively. Matrices are represented by uppercase letters in sans-serif font (e.g., $\mathsf{M}$) and vectors are in boldface font (e.g., $\mathbf{v}$). We use calligraphic symbols (e.g., $\mathcal{X}$) and the Greek letter $\Omega$ to denote sets. The probability distribution of a random variable $X$ is denoted by $p_X$. Denote by $|\cdot|$ the cardinality of a set and by $\mathbbold{1}\{\cdot\}$ the indicator function of an event. We denote $[a] := \{1,2,\cdots, \lfloor a \rfloor\}$ for all $a\ge 1$, $X^k := (X_1,X_2,\cdots,X_k)$, and $X(\Omega) = (X_i:i\in\Omega)$. Throughout the paper, all logarithms are to the base two.

The usual notation for entropy, $H(X)$, and mutual information, $I(X;Y)$, is used. We follow the $\epsilon$--$\delta$ notation in \cite{ElGamal:11} and the robust typicality introduced in \cite{Orlitsky:01}: For $X\sim p_X$ and $\epsilon\in(0,1)$, the set of typical sequences of length $k$ with respect to the probability distribution $p_X$ and the parameter $\epsilon$ is denoted by $\mathcal{T}_\epsilon^{(k)}(X)$, which is defined as 
\begin{IEEEeqnarray*}{rCl}
\mathcal{T}_\epsilon^{(k)}(X) := \left\{x^k\in\mathcal{X}^k :\left|\frac{\#(a|x^k)}{k}-p_X(a)\right|\le \epsilon p_X(a), \forall a\in \mathcal{X}\right\}, 
\end{IEEEeqnarray*}
where $\#(a|x^k)$ is the number of occurrences of $a$ in $x^k$.
Finally, the total correlation among the random variables $X(\Omega)$ is defined as 
\begin{IEEEeqnarray*}{rCl}
\Gamma(X(\Omega)) &:=& \sum_{i\in\Omega} H(X_i) - H(X(\Omega)).
\end{IEEEeqnarray*}

\section{Problem Statement} \label{sec:prob_state}
\begin{figure}[t!]
\begin{center}
\includegraphics[scale=0.8, trim={1.7cm 13.5cm 4cm 11.1cm}, clip]{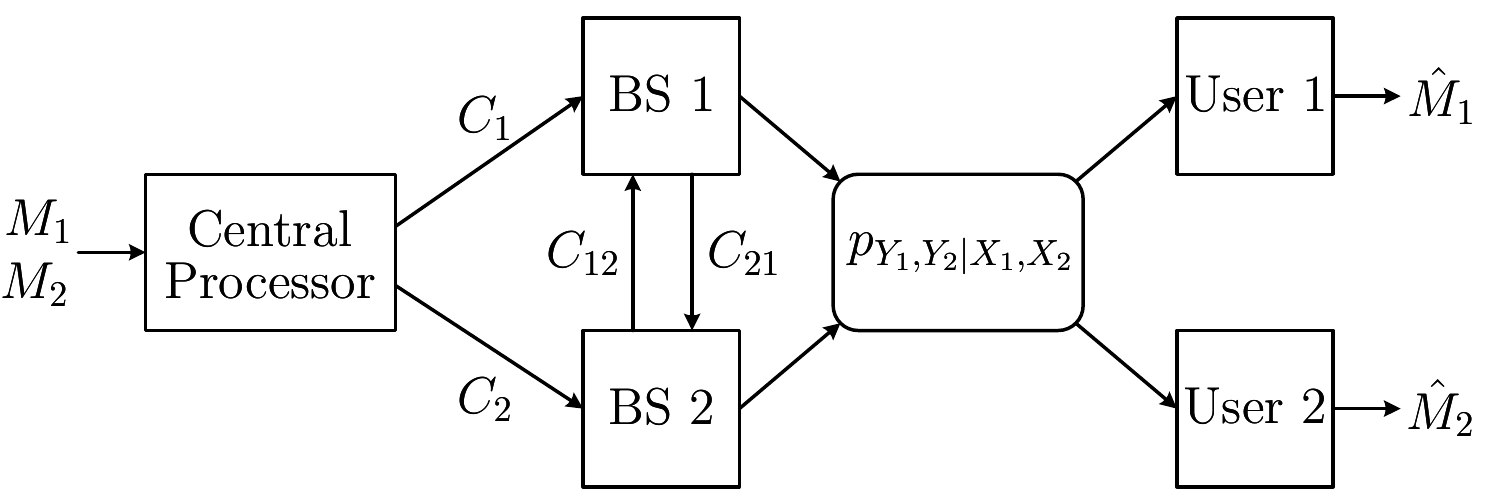}
\end{center}
\vspace{-0.15in}
\caption{Downlink C-RAN with BS cooperation: $2$ base stations and $2$ mobile users.}
\label{fig:system_coop}
\vspace{-0.1in}
\end{figure}

Consider the downlink $2$-BS $2$-user C-RAN with BS cooperation depicted in Figure~\ref{fig:system_coop}. The network consists of one central processor, two BSs, and two mobile users. The central processor communicates with the two BSs through individual noiseless bit pipes of finite capacities. Denote by $C_k$ the capacity of the link from the central processor to BS $k$. In addition, the two BSs can also communicate with each other through individual noiseless bit pipes of finite capacities. Denote by $C_{kj}$ the capacity of the link from BS $j$ to BS $k$. The network from the BSs to the mobile users is modeled as a discrete memoryless interference channel (DM-IC) $\langle \mathcal{X}_1\times\mathcal{X}_2, p_{Y_1,Y_2|X_1,X_2},\mathcal{Y}_1\times\mathcal{Y}_2\rangle$ that consists of four finite sets $\mathcal{X}_1,\mathcal{X}_2,\mathcal{Y}_1,\mathcal{Y}_2$ and a collection of conditional probability mass functions (pmf) $p_{Y_1,Y_2|X_1,X_2}$. 

With the help of the two BSs, the central processor wants to communicate two messages $M_1$ and $M_2$ to users~1 and~2, respectively. Assume that $M_1$ and $M_2$ are independent and uniformly distributed over $[2^{nR_1}]$ and $[2^{nR_2}]$, respectively. In this paper, we restrict attention to information processing on a block-by-block basis. Each block consists of a sequence of $n$ symbols. The entire communication is divided into three successive phases:
\begin{enumerate}[leftmargin=*]
\item central processor to BSs \\
The central processor conveys two indices $(W_1,W_2) := f_0(M_1,M_2)$
to BS~1 and BS~2, respectively, where $f_0:[2^{nR_1}]\times[2^{nR_2}]\to[2^{nC_1}]\times[2^{nC_2}]$ is the encoder of the central processor.
\item BS to BS conferencing communication \\
BS~1 conveys an index $W_{21} := f_1(W_1)$ to BS~2, where $f_1:[2^{nC_1}]\to[2^{nC_{21}}]$ is the conferencing encoder of BS~1. BS~2 conveys an index 
$W_{12} := f_2(W_2)$ to BS~1, where $f_2:[2^{nC_2}]\to[2^{nC_{12}}]$ is the conferencing encoder of BS~2. 
\item BSs to mobile users \\
BS~1 transmits a sequence $X_1^n := g_1(W_1,W_{12})$ over the DM-IC, where $g_1:[2^{nC_1}]\times[2^{nC_{12}}]\to\mathcal{X}_1^n$ is the channel encoder of BS~1. BS~2 transmits a sequence $X_2^n := g_2(W_2,W_{21})$ over the DM-IC, where $g_2:[2^{nC_2}]\times[2^{nC_{21}}]\to\mathcal{X}_2^n$ is the channel encoder of BS~2.
\end{enumerate}
Upon receiving the sequence $Y_\ell^n\in\mathcal{Y}_\ell^n$, user~$\ell\in\{1,2\}$ finds an estimate $\hat{M}_\ell := d_\ell(Y_\ell^n)$ of message $M_\ell$, where $d_\ell:\mathcal{Y}_\ell^n\to[2^{nR_\ell}]$ is the decoder of user~$\ell$.
The collection of the encoders $f_0,f_1,f_2,g_1,g_2$ and the decoders $d_1,d_2$ is called a $(2^{nR_1},2^{nR_2},n)$ channel code for the downlink $2$-BS $2$-user C-RAN model with BS cooperation. 

The average error probability is defined as 
\begin{IEEEeqnarray*}{rCl}
{\sf P}_e^{(n)} &:=& \mathbb{P}\left(\bigcup_{\ell=1}^2\{\hat{M}_\ell \neq M_\ell\}\right).
\end{IEEEeqnarray*}
We say that a rate pair $(R_1,R_2)$ is achievable if there exists a sequence of $(2^{nR_1},2^{nR_2},n)$ codes such that $\lim_{n\to \infty} {\sf P}_e^{(n)}=0$. The capacity region of the downlink C-RAN is the closure of the set of achievable rate pairs. 

Finally, we remark that using the discretization procedure \cite[Section 3.4.1]{ElGamal:11} and appropriately introducing input costs, our developed results for DM-ICs can be adapted to the Gaussian interference channel with constrained input power. The input--output relation of this channel is 
\begin{IEEEeqnarray}{rCl} \label{eq:Gaussian_model}
\begin{bmatrix} Y_1 \\ Y_2 \end{bmatrix} &=& \begin{bmatrix}  g_{11} & g_{12} \\ g_{21} & g_{22} \end{bmatrix}  \begin{bmatrix} X_1 \\ X_2 \end{bmatrix} + \begin{bmatrix} Z_1 \\ Z_2 \end{bmatrix}, 
\end{IEEEeqnarray}
where $X_k\in\mathbb{R}$ is the channel input from BS $k$, $Y_\ell$ is the channel output observed at user $\ell$, $g_{\ell k}\in\mathbb{R}$ is the channel gain from BS $k$ to user $\ell$, and $(Z_1,Z_2)$ are i.i.d. $\mathcal{N}(0,1)$ and each BS has to satisfy an average power constraint $P$, i.e., $\frac{1}{n}\sum_{i=1}^n x_{ki}^2 \le P$ for all $k\in\{1,2\}$.

\section{Generalized Data-Sharing Scheme} \label{sec:description}
We now propose a new coding scheme, which we term \emph{generalized data-sharing (G-DS) scheme} and which generalizes the data-sharing scheme \cite{Zakhour:11}. It is instructive to briefly review the encoding of the data-sharing scheme before presenting the details of our new G-DS scheme. 

\subsection{Preliminary: Data-Sharing Scheme} \label{subsec:prel}

The conventional data-sharing scheme follows from a {\em rate-splitting} approach: Each message $M_\ell$ is split into three independent submessages $M_{\ell0}$, $M_{\ell1}$, and $M_{\ell2}$, where $\ell\in\{1,2\}$. The central processor sends the private messages $(M_{1k},M_{2k})$ to BS $k$, where $k\in\{1,2\}$, and the common messages $(M_{10},M_{20})$ to both BSs. The BSs map the received submessages into codewords, i.e., $m_{1j} \to u_j^n$ and $m_{2j} \to v_j^n$, for all $j\in\{0,1,2\}$, and each BS $k\in\{1,2\}$ applies a symbol-by-symbol mapping $x_k(u_0,v_0,u_k,v_k)$ to map the codewords $(U_0^n,V_0^n,U_k^n,V_k^n)$ into channel inputs $X_k^n$. In the conventional data-sharing scheme, the codewords are generated according to the distribution  
\begin{IEEEeqnarray*}{rCl} 
p_{U_0,U_1,U_2,V_0,V_1,V_2} &=& \prod_{j=0}^2 p_{U_j}p_{V_j}.
\end{IEEEeqnarray*}
In \cite{Zakhour:11}, Zakhour and Gesbert specialized the data-sharing scheme to the memoryless Gaussian model and to linear mapping of $x_k$, which is known as {\em linear beamforming} in the literature.

Our aim is to develop a coding scheme that allows to exploit the full joint distribution $p_{U_0,U_1,U_2,V_0,V_1,V_2}$. However, to the best of our knowledge, the rate-splitting approach seems to admit at best the structure $p_{U_0,V_0}\prod_{j=1}^2p_{U_j,V_j|U_0,V_0}$. The reason is that since the private messages are independent of each other and of the common messages, the BSs cannot coordinate with each other to have $(U_1,V_1)$ directly correlate with $(U_2,V_2)$. In order to overcome this obstacle, we modify and extend the Liu--Kang scheme~\cite{Liu:14}: Each message, instead of being split into three independent parts, is now represented by a set of auxiliary index tuples. Each auxiliary index is referred to a codeword of independently generated codebooks. Through joint typicality test, we find auxiliary indices such that the set of corresponding codewords are coordinated. 

\subsection{Performance}
First, let us give a high-level summary of the G-DS scheme. The encoding is based on {\em multicoding}. We fix a joint pmf $p_{U_0,V_0,U_1,V_1,U_2,V_2}$ and independently generate six codebooks ${\sf U}_j$, ${\sf V}_j$, $j\in\{0,1,2\}$, from the marginals $p_{U_j}$, $p_{V_j}$, $j\in\{0,1,2\}$, respectively. For $j\in\{0,1,2\}$, the codebook ${\sf U}_j$ contains $2^{nR_{{\sf u}j}}$ codewords and the codebook ${\sf V}_j$ contains $2^{nR_{{\sf v}j}}$ codewords. Each message $m_1\in[2^{nR_1}]$ is associated with a unique bin $\mathcal{B}(m_1)$ of index tuples $(k_0,k_1,k_2)\in[2^{R_{{\sf u}0}}]\times[2^{R_{{\sf u}1}}]\times[2^{R_{{\sf u}2}}]$, which are indices to the codebooks ${\sf U}_0,{\sf U}_1,{\sf U}_2$, respectively. Similarly, each message $m_2\in[2^{nR_2}]$ is associated with a unique bin $\mathcal{B}(m_2)$ of index tuples $(\ell_0,\ell_1,\ell_2)\in[2^{R_{{\sf v}0}}]\times[2^{R_{{\sf v}1}}]\times[2^{R_{{\sf v}2}}]$, which are indices to the independently generated codebooks ${\sf V}_0,{\sf V}_1,{\sf V}_2$, respectively. Then, given $(m_1,m_2)$, we apply joint typicality encoding to find index tuples $(k_0,k_1,k_2)\in\mathcal{B}(m_1)$ and $(\ell_0,\ell_1,\ell_2)\in\mathcal{B}(m_2)$ such that $(U_0^n(k_0),U_1^n(k_1),U_2^n(k_2),V_0^n(\ell_0),V_1^n(\ell_1),V_2^n(\ell_2))$ are jointly typical. 

\begin{remark}
In addition to including the common auxiliaries $U_0$ and $V_0$, as already mentioned in \cite{Liu:14}, the main difference of our proposed scheme from the Liu--Kang scheme is that we do not enumerate the jointly typical pairs $(U_1^n(k_1),U_2^n(k_2))$ and $(V_1^n(\ell_1),V_2^n(\ell_2))$, which renders the analysis of the success probability of finding jointly typical tuples $(U_1^n(k_1),U_2^n(k_2),V_1^n(\ell_1),V_2^n(\ell_2))$ difficult. 
\hfill$\lozenge$\end{remark}

The next step is to convey $(k_0,\ell_0,k_1,\ell_1)$ to BS~1 and $(k_0,\ell_0,k_2,\ell_2)$ to BS~2. By taking advantage of the following facts, we can reduce the conventional sum rate $R_{{\sf u}0}+R_{{\sf v}0}+R_{{\sf u}j}+R_{{\sf v}j}$, $j\in\{1,2\}$:
\begin{enumerate}[leftmargin=*]
\item {\em Correlated index tuples} \\
The index tuple to be sent represents certain jointly typical codewords. As long as $U_0,V_0,U_j,V_j$ are not mutually independent, some members of $[2^{nR_{{\sf u}0}}]\times[2^{nR_{{\sf v}0}}]\times[2^{nR_{{\sf u}j}}]\times[2^{nR_{{\sf v}j}}]$ will never be used. Thus, instead of sending $(k_0, \ell_0, k_j, \ell_j)$ separately, we can enumerate all jointly typical codewords and simply convey an enumeration index.
\item {\em Opportunity of exploiting the cooperation links} \\
In the presence of cooperation links, the BSs do not need to learn all the information directly over the link from the central processor, but can learn part of it over the cooperation link.
\end{enumerate}

Finally, user~1 applies joint typicality decoding to recover $(k_0,k_1,k_2)$ and then the message $m_1$ can be uniquely identified. Similarly, user~2 applies joint typicality decoding to recover $(\ell_0,\ell_1,\ell_2)$ and then the message $m_2$ can be uniquely identified.

The achieved rate region of the G-DS scheme is presented in the following theorem. 

\begin{theorem} \label{thm:description}
A rate pair $(R_1,R_2)$ is achievable for the downlink $2$-BS $2$-user C-RAN with BS cooperation if there exist some rates $R_{{\sf u}j},R_{{\sf v}j}\ge 0$, $j\in\{0,1,2\}$, some joint pmf $p_{U_0,V_0,U_1,V_1,U_2,V_2}$, and some functions $x_k(u_0,v_0,u_k,v_k)$, $k\in\{1,2\}$, such that for all $\Omega_{\sf u},\Omega_{\sf v}\subseteq\{0,1,2\}$ satisfying $|\Omega_{\sf u}|+|\Omega_{\sf v}|\ge 2$,
\begin{IEEEeqnarray}{rCl}
\label{eq:Dcond1}
\mathbbold{1}\{|\Omega_{\sf u}|=3\}R_1 + \mathbbold{1}\{|\Omega_{\sf v}|=3\}R_2  &<& \sum_{i\in\Omega_{\sf u}} R_{{\sf u}i} + \sum_{j\in\Omega_{\sf v}} R_{{\sf v}j} - \Gamma(U(\Omega_{\sf u}),V(\Omega_{\sf v})); 
\end{IEEEeqnarray}
for all non-empty $\Omega_{\sf u},\Omega_{\sf v}\subseteq\{0,1,2\}$, 
\begin{IEEEeqnarray}{rCl}
\label{eq:Dcond2}
\sum_{i\in\Omega_{\sf u}} R_{{\sf u}i} &<& I(U(\Omega_{\sf u});U(\Omega_{\sf u}^c),Y_1) + \Gamma(U(\Omega_{\sf u})), \\
\label{eq:Dcond3}
\sum_{j\in\Omega_{\sf v}} R_{{\sf v}j} &<& I(V(\Omega_{\sf v});V(\Omega_{\sf v}^c),Y_2) + \Gamma(V(\Omega_{\sf v})); 
\end{IEEEeqnarray}
and 
\begin{IEEEeqnarray}{rCl}
\label{eq:Dcond4}
\sum_{i=\{0,1\}} R_{{\sf u}i} + \sum_{j\in\{0,1\}} R_{{\sf v}j} &<& C_1 + C_{12} + \Gamma(U_0,V_0,U_1,V_1), \\
\label{eq:Dcond5}
\sum_{i\in\{0,2\}} R_{{\sf u}i} + \sum_{j\in\{0,2\}} R_{{\sf v}j} &<& C_2 + C_{21} + \Gamma(U_0,V_0,U_2,V_2), \\
\label{eq:Dcond6}
\sum_{i=0}^2 R_{{\sf u}i} + \sum_{j=0}^2 R_{{\sf v}j} &<& C_1 + C_2 + \Gamma(U_0,V_0,U_1,V_1) +\Gamma(U_0,V_0,U_2,V_2) - \Gamma(U_0,V_0).
\end{IEEEeqnarray}
\end{theorem}

Unfortunately, the rate region in Theorem~\ref{thm:description} is hard to evaluate. Besides, we find it insightful to learn the effects of different code components. Thus, now we present three corollaries to Theorem~\ref{thm:description} where we restrict the correlation structure: 
\begin{enumerate}
\item Corollary \ref{col:dsI}: $U_j=V_j=\emptyset$ and $R_{{\sf u}j}=R_{{\sf v}j}=0$, $j\in\{1,2\}$,
\item Corollary \ref{col:dsII}: $p_{U_0,V_0,U_1,V_1,U_2,V_2}=\prod_{j=1}^2p_{U_j}p_{V_j}$,
\item Corollary \ref{col:dsIII}: $U_0=V_0=\emptyset$ and $R_{{\sf u}0}=R_{{\sf v}0}=0$.
\end{enumerate}
In all the corollaries, the auxiliaries $(R_{{\sf u}j},R_{{\sf v}j}:j\in\{0,1,2\})$ are eliminated through the Fourier--Motzkin elimination.\footnote{In this paper, all Fourier--Motzkin eliminations are performed using the software developed by Gattegno, {\em et al.}\cite{Gattegno:16}.} We remark that the first two correlation structures can also be realized through the rate-splitting approach mentioned in Section~\ref{subsec:prel}.

\begin{corollary}[Scheme I] \label{col:dsI}
A rate pair $(R_1, R_2)$ is achievable for the downlink $2$-BS $2$-user C-RAN with BS cooperation if
\begin{IEEEeqnarray*}{rCl}
R_1 &<& I(U_0;Y_1), \\
R_2 &<& I(V_0;Y_2), \\
R_1 + R_2 &<& I(U_0;Y_1) + I(V_0;Y_2) - I(U_0;V_0), \\
R_1 + R_2 &<& \min\{C_1+C_{12}, C_2+C_{21}, C_1+C_2\},
\end{IEEEeqnarray*}
for some joint pmf $p_{U_0,V_0}$ and some functions $x_k(u_0,v_0)$, $k\in\{1,2\}$.
\end{corollary}

\begin{corollary}[Scheme II] \label{col:dsII}
A rate pair $(R_1, R_2)$ is achievable for the downlink $2$-BS $2$-user C-RAN with BS cooperation if
\begin{IEEEeqnarray*}{rCl}
R_1 &<& C_1+C_{12}+I(U_2;Y_1|U_0,U_1), \\
R_1 &<& C_2+C_{21}+I(U_1;Y_1|U_0,U_2), \\
R_1 &<& I(U_0,U_1,U_2;Y_1), \\
R_2 &<& C_1+C_{12}+I(V_2;Y_2|V_0,V_1), \\
R_2 &<& C_2+C_{21}+I(V_1;Y_2|V_0,V_2), \\
R_2 &<& I(V_0,V_1,V_2;Y_2), \\
R_1+R_2 &<& C_1+C_2, \\
R_1+R_2 &<& C_1+C_{12}+I(U_2;Y_1|U_0,U_1)+I(V_2;Y_2|V_0,V_1), \\
R_1+R_2 &<& C_2+C_{21}+I(U_1;Y_1|U_0,U_2)+I(V_1;Y_2|V_0,V_2), \\
R_1+2R_2 &<& C_1+C_2+C_{12}+C_{21}+I(V_1,V_2;Y_2|V_0), \\
2R_1+R_2 &<& C_1+C_2+C_{12}+C_{21}+I(U_1,U_2;Y_1|U_0), \\
2R_1+2R_2 &<& C_1+C_2+C_{12}+C_{21}+I(U_1,U_2;Y_1|U_0)+I(V_1,V_2;Y_2|V_0), 
\end{IEEEeqnarray*}
for some joint pmf $\prod_{j=0}^2p_{U_j}p_{V_j}$ and some functions $x_k(u_0,v_0,u_k,v_k)$, $k\in\{1,2\}$.
\end{corollary}

When applied to the memoryless Gaussian model~\eqref{eq:Gaussian_model}, Corollary~\ref{col:dsII} with $C_{12}=C_{21}=0$ recovers the rate region of the scheme of Zakhour and Gesbert~\cite[Proposition~1]{Zakhour:11}.

\begin{corollary}[Scheme III] \label{col:dsIII}
A rate pair $(R_1, R_2)$ is achievable for the downlink $2$-BS $2$-user C-RAN with BS cooperation if
\begin{IEEEeqnarray*}{rCl}
R_1 &<& C_1+C_{12} + I(U_2;U_1,Y_1) - I(U_2;U_1,V_1), \\
R_1 &<& C_2+C_{21} + I(U_1;U_2,Y_1) - I(U_1;U_2,V_2), \\
R_1 &<& I(U_1,U_2;Y_1) + \min\left\{\begin{array}{l} 0, \\
I(V_1;V_2,Y_2) - I(V_1;U_1,U_2), \\
I(V_2;V_1,Y_2) - I(V_2;U_1,U_2) \end{array}\right\}, \\
R_2 &<& C_1+C_{12}+I(V_2;V_1,Y_2)-I(V_2;U_1,V_1), \\
R_2 &<& C_2+C_{21}+I(V_1;V_2,Y_2)-I(V_1;U_2,V_2), \\
R_2 &<& I(V_1,V_2;Y_2) + \min\left\{\begin{array}{l} 0, \\
I(U_2;U_1,Y_1) - I(U_2;V_1,V_2), \\
I(U_1;U_2,Y_1) - I(U_1;V_1,V_2) \end{array}\right\}, \\
R_1+R_2 &<& I(U_1,U_2;Y_1) + I(V_1,V_2;Y_2) - I(U_1,U_2;V_1,V_2), \\
R_1+R_2 &<& C_1+C_2 - I(U_1,V_1;U_2,V_2), \\
R_1+R_2 &<& C_1+C_{12} - I(U_1,V_1;U_2,V_2) \\
&& + \min\left\{\begin{array}{l} I(U_2;U_1,Y_1) + I(V_2;V_1,Y_2) - I(U_2;V_2), \\
2I(U_2;U_1,Y_1) + I(V_1,V_2;Y_2) - I(U_2;V_1) - I(U_2;V_2) + I(V_1;V_2), \\
I(U_1,U_2;Y_1) + 2I(V_2;V_1,Y_2) - I(U_1;V_2) - I(U_2;V_2) + I(U_1;U_2) \end{array}\right\}, \\
R_1+R_2 &<& C_2+C_{21} - I(U_1,V_1;U_2,V_2) \\
&& + \min\left\{\begin{array}{l} I(U_1;U_2,Y_1) + I(V_1;V_2,Y_2) - I(U_1;V_1), \\
2I(U_1;U_2,Y_1) + I(V_1,V_2;Y_2) - I(U_1;V_1) - I(U_1;V_2) + I(V_1;V_2) , \\
I(U_1,U_2;Y_1) + 2I(V_1;V_2,Y_2) - I(U_1;V_1) - I(U_2;V_1) + I(U_1;U_2) \end{array}\right\},
\end{IEEEeqnarray*}
for some joint pmf $p_{U_1,V_1,U_2,V_2}$ and some functions $x_k(u_k,v_k)$, $k\in\{1,2\}$ such that 
\begin{IEEEeqnarray*}{rCl}
I(U_1;V_1) &<& I(U_1;U_2,Y_1) + I(V_1;V_2,Y_2), \\
I(U_2;V_2) &<& I(U_2;U_1,Y_1) + I(V_2;V_1,Y_2), \\
I(U_1;V_2) &<& I(U_1;U_2,Y_1) + I(V_2;V_1,Y_2), \\
I(U_2;V_1) &<& I(U_2;U_1,Y_1) + I(V_1;V_2,Y_2).
\end{IEEEeqnarray*}
\end{corollary}

\subsection{Examples}
Now let us consider two special cases with simpler topologies. 
\begin{example}[$1$ BS and $2$ users]
The downlink $1$-BS $2$-user C-RAN can be considered as a special case of the downlink $2$-BS $2$-user C-RAN with $C_2=C_{12}=C_{21}=0$ and $p_{Y_1,Y_2|X_1,X_2}=p_{Y_1,Y_2|X_1}$. We fix a joint pmf $p_{U,V}$ and substitute $(U_1,V_1)=(U,V)$, $U_j=V_j=\emptyset$, and $R_{{\sf u}j}=R_{{\sf v}j}=0$, $j\in\{0,2\}$, in Theorem \ref{thm:description}. Then, after removing $R_{{\sf u}1}$ and $R_{{\sf v}1}$ by the Fourier--Motzkin elimination, we have the following corollary.
\begin{corollary}\label{col:1b2u}
A rate pair $(R_1,R_2)$ is achievable for the downlink $1$-BS $2$-user C-RAN if there exist some pmf $p_{U,V}$ and some function $x_1(u,v)$ such that 
\begin{IEEEeqnarray*}{rCl}
R_1 &<& I(U;Y_1), \\
R_2 &<& I(V;Y_2), \\
R_1+R_2 &<& I(U;Y_1) + I(V;Y_2) - I(U;V), \\
\label{eq:cond_dig}
R_1+R_2 &<& C_1. \IEEEyesnumber
\end{IEEEeqnarray*}
\end{corollary}
Thus, the achieved rate region is essentially Marton's inner bound~\cite{Marton:79} with the additional constraint \eqref{eq:cond_dig} due to the fact that the digital link is of finite capacity.
\hfill$\lozenge$\end{example}

\begin{example}[$2$ BSs and $1$ user]
The downlink $2$-BS $1$-user C-RAN is a class of {\em diamond networks} \cite{Kang:15, SaeediBidokhti:15}, which can be considered as a special case of the downlink $2$-BS $2$-user C-RAN by setting $R_2=0$. We fix a joint pmf $p_{U,X_1,X_2}$ and substitute $(U_0,U_1,U_2)=(U,X_1,X_2)$, $V_j=\emptyset$, and $R_{{\sf v}j}=0$, $j\in\{0,1,2\}$, in Theorem \ref{thm:description}. Then, after removing $R_{{\sf u}0}$, $R_{{\sf u}1}$, and $R_{{\sf u}2}$ by the Fourier--Motzkin elimination, we have the following corollary.
\begin{corollary}\label{col:2b1u}
Any rate $R_1$ is achievable for the downlink $2$-BS $1$-user C-RAN with BS cooperation if there exists some pmf $p_{U,X_1,X_2}$ such that 
\begin{IEEEeqnarray*}{rCl} 
R_1 &<& \min\left\{\begin{array}{l}C_1+C_2-I(X_1;X_2|U), \\
C_1+C_{12}+I(X_2;Y_1|U,X_1), \\
C_2+C_{21}+I(X_1;Y_1|U,X_2), \\
I(X_1,X_2;Y_1), \\
\frac{1}{2}[C_1+C_2+C_{12}+C_{21}+I(X_1,X_2;Y_1|U)-I(X_1;X_2|U)]
\end{array} \right\}.
\end{IEEEeqnarray*}
\end{corollary}

\begin{remark}
Considering diamond networks with an orthogonal broadcast channel, the proposed G-DS scheme recovers the achievability results in \cite[Theorem 2]{Kang:15} and \cite[Theorem 1]{SaeediBidokhti:15}. It is shown in \cite{SaeediBidokhti:15} that the achievability is optimal when the second hop is the binary-adder multiple-access channel, i.e., $\mathcal{X}_1=\mathcal{X}_2=\{0,1\}$, $\mathcal{Y}_1=\{0,1,2\}$, and $Y_1=X_1+X_2$. Furthermore, the proposed G-DS scheme
recovers the achievability result in \cite[Theorem 2]{Zhao:15} in which cooperation between relays is also included in the network model.
\end{remark}
\hfill$\lozenge$\end{example}

\subsection{Coding Scheme}
{\em Codebook generation:} Fix a joint pmf $p_{U_0,V_0,U_1,V_1,U_2,V_2}$ and functions $x_j(u_0,v_0,u_j,v_j)$, $j\in\{1,2\}$. Randomly and independently generate sequences
\begin{itemize}[leftmargin=*]
\item $u_0^n(k_0)$, each according to $\prod_{i=1}^np_{U_0}(u_{0i})$, for $k_0\in [2^{nR_{{\sf u}_0}}]$; 
\item $u_1^n(k_1)$, each according to $\prod_{i=1}^np_{U_1}(u_{1i})$, for $k_1\in[2^{nR_{{\sf u}_1}}]$;
\item $u_2^n(k_2)$, each according to $\prod_{i=1}^np_{U_2}(u_{2i})$, for $k_2\in[2^{nR_{{\sf u}_2}}]$;
\item $v_0^n(\ell_0)$, each according to $\prod_{i=1}^np_{V_0}(v_{0i})$, for $\ell_0 \in [2^{nR_{{\sf v}_0}}]$;
\item $v_1^n(\ell_1)$, each according to $\prod_{i=1}^np_{V_1}(v_{1i})$, for $\ell_1\in [2^{nR_{{\sf v}_1}}]$;
\item $v_2^n(\ell_2)$, each according to $\prod_{i=1}^np_{V_2}(v_{2i})$, for $\ell_2\in [2^{nR_{{\sf v}_2}}]$. 
\end{itemize}
Next, we generate three dictionaries: 
\begin{IEEEeqnarray*}{rCl}
\mathcal{D}_0 &=& \{(k_0,\ell_0)\in[2^{nR_{{\sf u}0}}]\times[2^{nR_{{\sf v}0}}]: (u_0^n(k_0),v_0^n(\ell_0))\in\mathcal{T}_{\epsilon'}^{(n)}\}, \\
\mathcal{D}_1(k,\ell) &=& \{(k_1,\ell_1)\in[2^{nR_{{\sf u}1}}]\times[2^{nR_{{\sf v}1}}]: (u_1^n(k_1),v_1^n(\ell_1))\in\mathcal{T}_{\epsilon'}^{(n)}(U_1,V_1|u_0^n(k),v_0^n(\ell))\}, \\
\mathcal{D}_2(k,\ell) &=& \{(k_2,\ell_2)\in[2^{nR_{{\sf u}2}}]\times[2^{nR_{{\sf v}2}}]: (u_2^n(k_2),v_2^n(\ell_2))\in\mathcal{T}_{\epsilon'}^{(n)}(U_2,V_2|u_0^n(k),v_0^n(\ell))\}, 
\end{IEEEeqnarray*}
for all $(k,\ell)\in\mathcal{D}_0$. Every index tuple in the dictionaries is assigned a unique reference label. For example, the first index tuple in $\mathcal{D}_1(k,\ell)$ is referred to as $\mathcal{D}_1(1|k,\ell)$. 
Also, we denote by $\mathcal{D}_j^{-1}$ the inverse map of $\mathcal{D}_j$, $j\in\{0,1,2\}$.

Finally, we randomly and independently assign an index $m_1(k_0,k_1,k_2)$ to each index tuple $(k_0,k_1,k_2)\in[2^{nR_{{\sf u}0}}]\times[2^{nR_{{\sf u}1}}]\times[2^{nR_{{\sf u}2}}]$ according to a uniform pmf over $[2^{nR_1}]$. Similarly, we randomly and independently assign an index $m_2(\ell_0,\ell_1,\ell_2)$ to each index tuple $(\ell_0,\ell_1,\ell_2)\in[2^{nR_{{\sf v}0}}]\times[2^{nR_{{\sf v}1}}]\times[2^{nR_{{\sf v}2}}]$ according to a uniform pmf over $[2^{nR_2}]$. We refer to each subset of index tuples with the same index $m_j$ as a bin $\mathcal{B}_j(m_j)$, $j\in\{1,2\}$.

\begin{figure}[t!]
\begin{center}
\includegraphics[scale=0.8, trim={2.5cm 2.8cm 5cm 16.5cm}, clip]{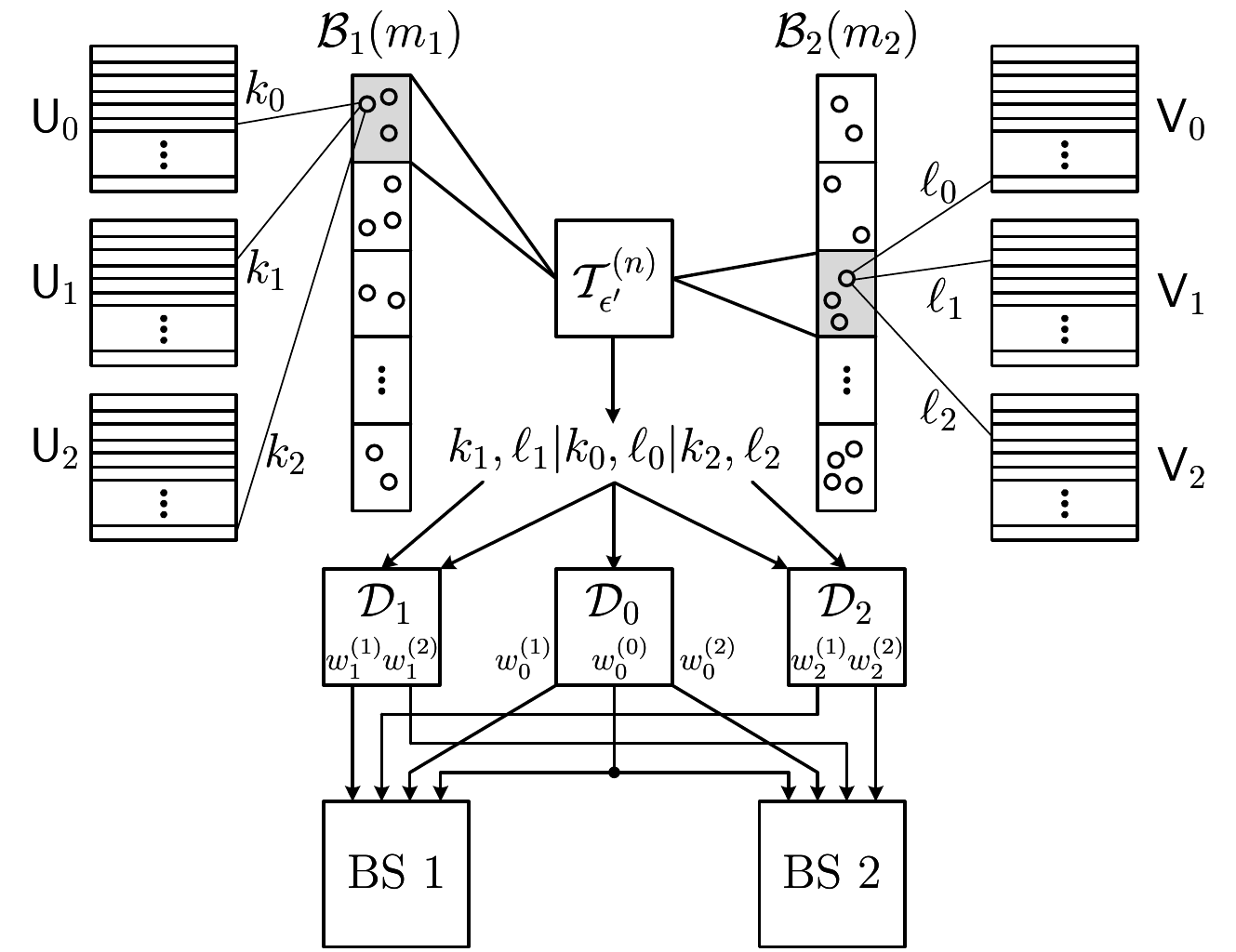}
\end{center}
\vspace{-0.2in}
\caption{Illustration of the encoding operation at the central processor in the G-DS scheme.}
\label{fig:description}
\vspace{-0.1in}
\end{figure}

{\em Central Processor:} Upon seeing $(m_1,m_2)$, the central processor finds 
$(k_0,k_1,k_2)\in\mathcal{B}_1(m_1)$ and $(\ell_0,\ell_1,\ell_2)\in\mathcal{B}_2(m_2)$ such that 
\begin{IEEEeqnarray*}{rCl}
(u_0^n(k_0),u_1^n(k_1),u_2^n(k_2),v_0^n(\ell_0),v_1^n(\ell_1),v_2^n(\ell_2))\in\mathcal{T}_{\epsilon'}^{(n)}.
\end{IEEEeqnarray*}
If there is more than one such tuple, choose an arbitrary one among them. If no such tuple exists, choose $(k_0,k_1,k_2,\ell_0,\ell_1,\ell_2)=(1,1,1,1,1,1)$. Then, the central processor splits $\mathcal{D}_0^{-1}(k_0,\ell_0)$ into three subindices $w_0^{(0)}$, $w_0^{(1)}$, and $w_0^{(2)}$ of rates $R_{00}$, $R_{01}$, and $R_{02}$, respectively. 
Also, for $j\in\{1,2\}$, the central processor splits the index $\mathcal{D}_j^{-1}(k_j,\ell_j|k_0,\ell_0)$ into two subindices $w_j^{(1)}$ and $w_j^{(2)}$ of rates $R_{j1}$ and $R_{j2}$, respectively. 
Finally, the central processor sends the index tuple $(w_0^{(0)}, w_0^{(1)},w_1^{(1)},w_2^{(1)})$ to BS~1 and $(w_0^{(0)}, w_0^{(2)},w_1^{(2)}, w_2^{(2)})$ to BS~2.
The encoding operation at the central processor is illustrated in Figure~\ref{fig:description}. 

{\em BS:} BS~1 forwards $(w_0^{(1)},w_2^{(1)})$ to BS~2 over the cooperation link. BS~2 forwards $(w_0^{(2)},w_1^{(2)})$ to BS~1 over the cooperation link. Given $(w_0^{(0)}, w_0^{(1)}, w_0^{(2)})$, both BSs can recover $\mathcal{D}_0^{-1}(k_0,\ell_0)$ and thus the common indices $(k_0,\ell_0)$. Then, BS~$j\in\{1,2\}$,  can recover $\mathcal{D}_j^{-1}(k_j,\ell_j|k_0,\ell_0)$ from $(w_j^{(1)}, w_j^{(2)})$ and $(k_0,\ell_0)$. Finally, BS~$j$ transmits the symbol $x_{ji}(u_{0i}(k_0),v_{0i}(\ell_0),u_{ji}(k_j),v_{ji}(\ell_j))$ at time $i\in[n]$. 

{\em Decoding:} Let $\epsilon>\epsilon'$. User~1 declares that $\hat{m}_1$ is sent if it is the unique message such that for some $(k_0,k_1,k_2)\in\mathcal{B}_1(\hat{m}_1)$ it holds that $(u_0^n(k_0),u_1^n(k_1),u_2^n(k_2),y_1^n)\in\mathcal{T}_\epsilon^{(n)}$; otherwise it declares an error. Decoder $2$ declares that $\hat{m}_2$ is sent if it is the unique message such that $(v_0^n(\ell_0),v_1^n(\ell_1),v_2^n(\ell_2),y_2^n)\in\mathcal{T}_\epsilon^{(n)}$ for some $(\ell_0,\ell_1,\ell_2)\in\mathcal{B}_2(\hat{m}_2)$; otherwise it declares an error.

{\em Analysis of Error Probability:}
Let $(M_1,M_2)$ be the messages and let $(K_0,K_1,K_2,L_0,L_1,L_2)$ be the indices chosen at the encoder. In order to have a lossless transmission over the digital links, it requires that 
\begin{IEEEeqnarray*}{rCl}
R_{00} + R_{01} + R_{11} + R_{21} &\le& C_1, \\
R_{00} + R_{02} + R_{12} + R_{22} &\le& C_2, \\
R_{01} + R_{21} &\le& C_{21}, \\
R_{02} + R_{12} &\le& C_{12}.
\end{IEEEeqnarray*}
Also, we note that 
\begin{IEEEeqnarray*}{rCl}
R_{00} + R_{01} + R_{02} &=& \log |\mathcal{D}_0|, \\
R_{11} + R_{12} &=& \log |\mathcal{D}_1(K_0,L_0)|, \\
R_{21} + R_{22} &=& \log |\mathcal{D}_2(K_0,L_0)|.
\end{IEEEeqnarray*}
Thus, after applying Fourier-Motzkin elimination to remove $R_{00}$ and $(R_{j1},R_{j2})$, $j\in\{0,1,2\}$, we have 
\begin{IEEEeqnarray}{rCl}
\label{eq:random1}
\log |\mathcal{D}_0| + \log |\mathcal{D}_1(K_0,L_0)|  &\le& C_1 + C_{12}, \\
\label{eq:random2}
\log |\mathcal{D}_0| + \log |\mathcal{D}_2(K_0,L_0)|  &\le& C_2 + C_{21}, \\
\label{eq:random3}
\log |\mathcal{D}_0| + \log |\mathcal{D}_1(K_0,L_0)| + \log |\mathcal{D}_2(K_0,L_0)|  &\le& C_1 + C_2. 
\end{IEEEeqnarray}
We denote by $\mathcal{A}$ the intersection of the random events~\eqref{eq:random1},~\eqref{eq:random2}, and~\eqref{eq:random3}. From Lemma \ref{lma:ind} proved in Appendix~\ref{apdx:ind}, the random event $\mathcal{A}$ happens with high probability as $n\to\infty$ if 
\begin{IEEEeqnarray*}{rCl}
C_1 + C_{12} &\ge& R_{{\sf u}0}+R_{{\sf v}0} - I(U_0;V_0) + R_{{\sf u}1}+R_{{\sf v}1} - I(U_1;V_1) - I(U_0,V_0;U_1,V_1), \\
C_2 + C_{21} &\ge& R_{{\sf u}0}+R_{{\sf v}0} - I(U_0;V_0) + R_{{\sf u}2}+R_{{\sf v}2} - I(U_2;V_2) - I(U_0,V_0;U_2,V_2), \\
C_1 + C_2 &\ge& R_{{\sf u}0}+R_{{\sf v}0} - I(U_0;V_0) + R_{{\sf u}1}+R_{{\sf v}1} - I(U_1;V_1) - I(U_0,V_0;U_1,V_1) \\
&& \hspace{3.75cm} + R_{{\sf u}2}+R_{{\sf v}2} - I(U_2;V_2) - I(U_0,V_0;U_2,V_2). 
\end{IEEEeqnarray*}

Besides the error event $\mathcal{A}^c$, the decoding at User~1 fails if one or more of the following events occur:
\begin{IEEEeqnarray*}{rCl}
\mathcal{E}_{\sf s} &=& \{(U_0^n(k_0),U_1^n(k_1),U_2^n(k_2),V_0^n(\ell_0),V_1^n(\ell_1),V_2^n(\ell_2))\notin\mathcal{T}_{\epsilon'}^{(n)} \\ && \hspace{0.5cm}\text{ for all } (k_0,k_1,k_2)\in\mathcal{B}_1(M_1),(\ell_0,\ell_1,\ell_2)\in\mathcal{B}_2(M_2)\}, \\
\mathcal{E}_{\sf d0} &=& \{(U_0^n(K_0),U_1^n(K_1),U_2^n(K_2),Y_1^n)\notin\mathcal{T}_\epsilon^{(n)} \}, \\
\mathcal{E}_{\sf d1} &=& \{(U_0(K_0),U_1^n(k_1),U_2^n(K_2),Y_1^n)\in\mathcal{T}_\epsilon^{(n)}\text{ for some } k_1\neq K_1 \}, \\
\mathcal{E}_{\sf d2} &=& \{(U_0(K_0),U_1^n(K_1),U_2^n(k_2),Y_1^n)\in\mathcal{T}_\epsilon^{(n)}\text{ for some } k_2\neq K_2 \}, \\
\mathcal{E}_{\sf d3} &=& \{(U_0(K_0),U_1^n(k_1),U_2^n(k_2),Y_1^n)\in\mathcal{T}_\epsilon^{(n)}\text{ for some } k_1\neq K_1, k_2\neq K_2 \} \\
\mathcal{E}_{\sf d4} &=& \{(U_0(k_0),U_1^n(K_1),U_2^n(K_2),Y_1^n)\in\mathcal{T}_\epsilon^{(n)}\text{ for some } k_0\neq K_0\}, \\
\mathcal{E}_{\sf d5} &=& \{(U_0(k_0),U_1^n(k_1),U_2^n(K_2),Y_1^n)\in\mathcal{T}_\epsilon^{(n)}\text{ for some } k_0\neq K_0, k_1\neq K_1 \}, \\
\mathcal{E}_{\sf d6} &=& \{(U_0(k_0),U_1^n(K_1),U_2^n(k_2),Y_1^n)\in\mathcal{T}_\epsilon^{(n)}\text{ for some } k_0\neq K_0, k_2\neq K_2 \}, \\
\mathcal{E}_{\sf d7} &=& \{(U_0(k_0),U_1^n(k_1),U_2^n(k_2),Y_1^n)\in\mathcal{T}_\epsilon^{(n)}\text{ for some } k_0\neq K_0, k_1\neq K_1, k_2\neq K_2 \}.
\end{IEEEeqnarray*}
Thus, the average error probability for $M_1$ is upper bounded as 
\begin{IEEEeqnarray*}{rCl}
\mathbb{P}(\{\hat{M_1}\neq M_1\}) &\le& \mathbb{P}(\mathcal{E}_{\sf s}) +  
\mathbb{P}(\mathcal{A}^c) + \mathbb{P}(\mathcal{E}_{\sf d0}\cap\mathcal{E}_{\sf s}^c\cap\mathcal{A})+\sum_{i=1}^{7}\mathbb{P}(\mathcal{E}_{di}).
\end{IEEEeqnarray*}

From Lemma \ref{lma:smcl} proved in Appendix~\ref{apdx:smcl}, the term $\mathbb{P}(\mathcal{E}_{\sf s})$ tends to zero as $n\to\infty$ if 
\begin{IEEEeqnarray*}{ll}
&\sum_{i\in\Omega_{\sf u}} R_{{\sf u}i} + \sum_{j\in\Omega_{\sf v}} R_{{\sf v}j} 
 \\
&> \mathbbold{1}\{\Omega_{\sf u}=\{0,1,2\}\}R_1 + \mathbbold{1}\{\Omega_{\sf v}=\{0,1,2\}\}R_2 + \Gamma(U(\Omega_{\sf u}),V(\Omega_{\sf v})), 
\end{IEEEeqnarray*}
for all $\Omega_{\sf u}, \Omega_{\sf v}\subseteq\{0,1,2\}$ such that $|\Omega_{\sf u}|+|\Omega_{\sf v}|\ge 2$.

Next, due to the codebook construction and the conditional typicality lemma~\cite[p. 27]{ElGamal:11}, $\mathbb{P}(\mathcal{E}_{\sf d0}\cap\mathcal{E}_{\sf s}^c\cap\mathcal{A})$ tends to zero as $n\to\infty$. Finally, using the joint typicality lemma~\cite[p. 29]{ElGamal:11}, $\sum_{i=1}^{7}\mathbb{P}(\mathcal{E}_{{\sf d}i})$ tends to zero as $n\to\infty$ if 
\begin{IEEEeqnarray*}{rCl}
R_{{\sf u}1}  &<& I(U_1;U_0,U_2,Y_1) - \delta(\epsilon), \\
R_{{\sf u}2}  &<& I(U_2;U_0,U_1,Y_1) - \delta(\epsilon), \\
R_{{\sf u}1}+R_{{\sf u}2}  &<& I(U_1,U_2;U_0,Y_1) + I(U_1;U_2) - \delta(\epsilon), \\
R_{{\sf u}0} &<& I(U_0;U_1,U_2,Y_1) - \delta(\epsilon), \\
R_{{\sf u}0}+R_{{\sf u}1}  &<& I(U_0,U_1;U_2,Y_1) + I(U_0;U_1)- \delta(\epsilon), \\
R_{{\sf u}0}+R_{{\sf u}2}  &<& I(U_0,U_2;U_1,Y_1) + I(U_0;U_2)- \delta(\epsilon), \\
R_{{\sf u}0}+R_{{\sf u}1}+R_{{\sf u}2}  &<& I(U_0,U_1,U_2;Y_1) + I(U_0;U_1,U_2) + I(U_1;U_2) - \delta(\epsilon).
\end{IEEEeqnarray*}

The average error probability for $M_2$ can be bounded in a similar manner and then we have the additional rate conditions
\begin{IEEEeqnarray*}{rCl}
R_{{\sf v}1}  &<& I(V_1;V_0,V_2,Y_2) - \delta(\epsilon), \\
R_{{\sf v}2}  &<& I(V_2;V_0,V_1,Y_2) - \delta(\epsilon), \\
R_{{\sf v}1}+R_{{\sf v}2}  &<& I(V_1,V_2;V_0,Y_2) + I(V_1;V_2) - \delta(\epsilon), \\
R_{{\sf v}0} &<& I(V_0;V_1,V_2,Y_2) - \delta(\epsilon), \\
R_{{\sf v}0}+R_{{\sf v}1}  &<& I(V_0,V_1;V_2,Y_2) + I(V_0;V_1)- \delta(\epsilon), \\
R_{{\sf v}0}+R_{{\sf v}2}  &<& I(V_0,V_2;V_1,Y_2) + I(V_0;V_2)- \delta(\epsilon), \\
R_{{\sf v}0}+R_{{\sf v}1}+R_{{\sf u}2}  &<& I(V_0,V_1,V_2;Y_2) + I(V_0;V_1,V_2) + I(V_1;V_2) - \delta(\epsilon).
\end{IEEEeqnarray*}
Finally, the theorem is established by letting $\epsilon$ tend to zero.


\section{Generalized Compression Scheme}\label{sec:compress}
This section is devoted to compression-based schemes, where our contributions are as follows: 
\begin{itemize}
\item For the downlink $2$-BS $2$-user C-RAN, we generalize the original compression scheme in~\cite{Park:13} in two directions: 1) We introduce a cloud center and incorporate BS cooperation; and 2) we derive a single-letter rate region for general discrete memoryless channels on the second hop. 

\item We simplify the achievable rate region of the DDF scheme for the general downlink $N$-BS $L$-user C-RAN with BS cooperation.
We provide a performance analysis of the DDF scheme under the memoryless Gaussian model and show that the DDF scheme achieves within a constant gap (independent of transmission power) from the capacity region.
\end{itemize}

\subsection{Performance and Coding Scheme}
We start with a high-level summary of the proposed G-Compression scheme. The encoding is based on superposition coding and multicoding. Each message $m_j$, $j\in\{1,2\}$, is associated with a set of independently generated codewords $U_j^n(m_j,\ell_j)$ of size $2^{n\tilde{R}_j}$. Then, we generate three codebooks ${\sf X}_0,{\sf X}_1,{\sf X}_2$ using superposition coding: the codebook ${\sf X}_0$ contains the cloud centers $X_0^n(k_0)$ and the codebooks ${\sf X}_1$ and ${\sf X}_2$ contain the satellite codewords $X_1^n(k_1|k_0)$ and $X_2^n(k_2|k_0)$, respectively.

Given $(m_1,m_2)$, we apply joint typicality encoding to find an index tuple $(k_0,k_1,k_2,\ell_1,\ell_2)$ such that $(U_1^n(m_1,\ell_1),$ $U_2^n(m_2,\ell_2),X_0^n(k_0),X_1^n(k_1),X_2^n(k_2))$ are jointly typical. In words, we first apply Marton's coding on the messages $m_1$ and $m_2$. Then, the auxiliaries $U_1^n(m_1,\ell_1)$ and $U_2^n(m_2,\ell_2)$ are compressed into three descriptions $X_0^n(k_0)$, $X_1^n(k_1|k_0)$, and $X_2^n(k_2|k_0)$.

The next step is to convey $(k_0,k_1)$ to BS~1 and $(k_0,k_2)$ to BS~2, during which the cooperation links are used to reduce the workload of the digital links from the central processor to the BSs. Finally, each user~$j\in\{1,2\}$ applies joint typicality decoding to recover the auxiliary $U_j^n(m_j,\ell_j)$ and thus can recover the desired message $m_j$.

\begin{theorem} \label{thm:compress}
A rate pair $(R_1, R_2)$ is achievable for the downlink $2$-BS $2$-user C-RAN with BS cooperation if 
\begin{IEEEeqnarray*}{rCl}
R_1 &<& I(U_1;Y_1) + \min\left\{\begin{array}{l}
0, \\
C_1+C_{12} - I(U_1;X_0,X_1), \\
C_2+C_{21} - I(U_1;X_0,X_2)
\end{array}\right\}, \\
R_2 &<& I(U_2;Y_2) + \min\left\{\begin{array}{l}
0, \\
C_1+C_{12} - I(U_2;X_0,X_1), \\
C_2+C_{21} - I(U_2;X_0,X_2)
\end{array}\right\}, \\
R_1+R_2 &<& I(U_1;Y_1)+I(U_2;Y_2)-I(U_1;U_2) \\
&& + \min\left\{\begin{array}{l}
0, \\
C_1+C_{12}-I(U_1,U_2;X_0,X_1), \\
C_2+C_{21} - I(U_1,U_2;X_0,X_2), \\
C_1+C_2-I(U_1,U_2;X_0,X_1,X_2)-I(X_1;X_2|X_0)
\end{array}\right\}, \\
2R_1+R_2 &<& I(U_1;Y_1)+I(U_2;Y_2)-I(U_1;U_2) \\
&& +C_1+C_2+C_{12}+C_{21}-I(U_1,U_2;X_0,X_1,X_2)-I(X_1;X_2|X_0), \\
&& +I(U_1;Y_1)-I(U_1;X_0), \\
R_1+2R_2 &<& I(U_1;Y_1)+I(U_2;Y_2)-I(U_1;U_2) \\
&& +C_1+C_2+C_{12}+C_{21}-I(U_1,U_2;X_0,X_1,X_2)-I(X_1;X_2|X_0), \\
&& +I(U_2;Y_2)-I(U_2;X_0), \\
2R_1+2R_2 &<& I(U_1;Y_1)+I(U_2;Y_2)-I(U_1;U_2) \\
&& +C_1+C_2+C_{12}+C_{21}-I(U_1,U_2;X_0,X_1,X_2)-I(X_1;X_2|X_0) \\
&& +I(U_1;Y_1)+I(U_2;Y_2)-I(U_1;U_2)-I(U_1,U_2;X_0),
\end{IEEEeqnarray*}
for some joint pmf $p_{U_1,U_2,X_0,X_1,X_2}$.
\end{theorem}

\begin{IEEEproof}
{\em Codebook generation:} Fix a joint pmf $p_{U_1,U_2,X_0,X_1,X_2}$. For $j\in\{1,2\}$, randomly and independently generate sequences $u_j^n(m_j,\ell_j)$, according to $\prod_{i=1}^np_{U_j}(u_{ji})$, for $(m_j,\ell_j)\in[2^{nR_j}]\times[2^{n\tilde{R}_j}]$. Randomly and independently generate sequences $x_0^n(k_0)$, according to $\prod_{i=1}^np_{X_0}(x_{0i})$, for $k_0\in[2^{nR'_0}]$. Finally, for $j\in\{1,2\}$, randomly and independently generate sequences $x_j^n(k_j|k_0)$, each according to $\prod_{i=1}^np_{X_j|X_0}(x_{ji}|x_{0i}(k_0))$, for $(k_0,k_j)\in[2^{nR'_0}]\times[2^{nR'_j}]$. 

\begin{figure}[t!]
\begin{center}
\includegraphics[scale=0.8, trim={0.5cm 0.5cm 9cm 19.5cm}, clip]{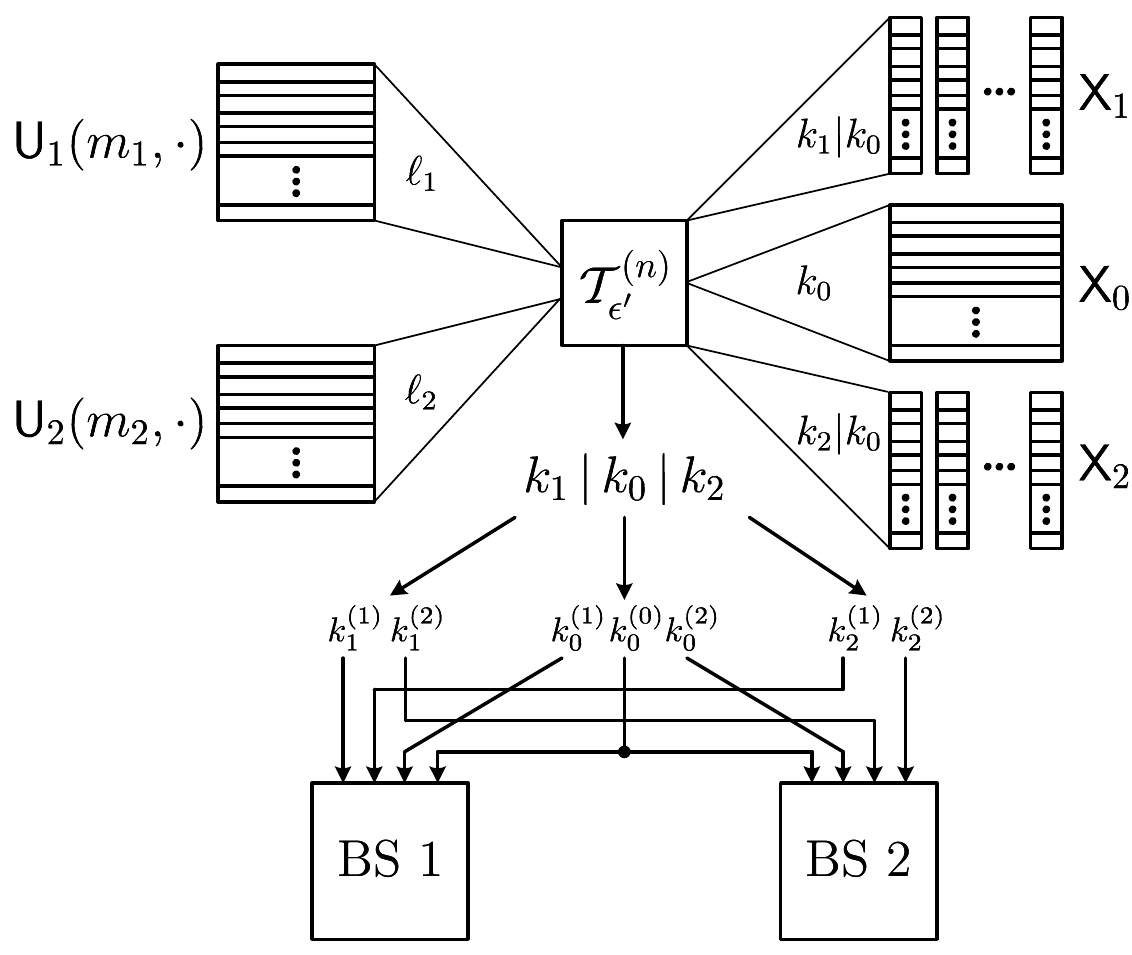}
\end{center}
\vspace{-0.2in}
\caption{Illustration of the encoding operation at the central processor in the G-Compression scheme.}
\label{fig:compression}
\vspace{-0.1in}
\end{figure}

{\em Central Processor:} Upon seeing $(m_1,m_2)$, the central processor finds an index tuple $(k_0,k_1,k_2,\ell_1,\ell_2)$ such that 
\begin{IEEEeqnarray*}{rCl}
(u_1^n(m_1,\ell_1),u_2^n(m_2,\ell_2),x_0^n(k_0),x_1^n(k_1|k_0),x_2^n(k_2|k_0))\in\mathcal{T}_{\epsilon'}^{(n)}.
\end{IEEEeqnarray*}
If there is more than one such tuple, choose an arbitrary one among them. If no such tuple exists, choose $(k_0,k_1,k_2,\ell_1,\ell_2)=(1,1,1,1,1)$. Then, the central processor splits $k_0$ into three subindices $k_0^{(0)}$, $k_0^{(1)}$, and $k_0^{(2)}$ of rates $R'_{00}$, $R'_{01}$, and $R'_{02}$, respectively. Also, for $j\in\{1,2\}$, the central processor splits $k_j$ into two subindices $k_j^{(1)}$ and $k_j^{(2)}$ of rates $R'_{j1}$ and $R'_{j2}$, respectively. Finally, the central processor sends the index tuple $(k_0^{(0)}, k_0^{(1)},k_1^{(1)},k_2^{(1)})$ to BS~1 and $(k_0^{(0)}, k_0^{(2)},k_1^{(2)},k_2^{(2)})$ to BS~2. The encoding operation at the central processor is illustrated in Figure~\ref{fig:compression}. 

{\em BS:} BS~1 forwards $(k_0^{(1)},k_2^{(1)})$ to BS~2 over the cooperation link. BS~2 forwards $(k_0^{(2)},k_1^{(2)})$ to BS~1 over the cooperation link. Thus, BS~$j\in\{1,2\}$ learns the value of $(k_0,k_j)$ and transmits $x_j^n(k_j|k_0)$.

{\em Decoding:} Let $\epsilon>\epsilon'$. For $j\in\{1,2\}$, upon seeing $y_j^n$, user~$j$ finds the unique pair $(\hat{m}_j,\hat{\ell}_j)$ such that $(u_j^n(\hat{m}_j,\hat{\ell}_j),y_j^n)\in\mathcal{T}_\epsilon^{(n)}$ and declares that $\hat{m}_j$ is sent; otherwise it declares an error.

{\em Analysis of Error Probability:}
Let $(M_1,M_2)$ be the messages and let $(K_0,K_1,K_2,L_1,L_2)$ be the indices chosen at the central processor. In order to have a lossless transmission over the digital links, it requires that 
\begin{IEEEeqnarray*}{rCl}
R'_{00} + R'_{01} + R'_{11} + R'_{21} &\le& C_1, \\
R'_{00} + R'_{02} + R'_{12} + R'_{22} &\le& C_2, \\
R'_{01} + R'_{21} &\le& C_{21}, \\
R'_{02} + R'_{12} &\le& C_{12}.
\end{IEEEeqnarray*}
Note that $R'_{00} + R'_{01} + R'_{02} = R'_0$ and  $R'_{j1} + R'_{j2} = R'_j$, $j\in\{1,2\}$. 

Assuming the above conditions are satisfied, the decoding at user~1 fails if one or more of the following events occur:
\begin{IEEEeqnarray*}{rCl}
\mathcal{E}_0 &=& \{(U_1^n(M_1,\ell_1),U_2^n(M_2,\ell_2),X_0^n(k_0),X_1^n(k_1|k_0),X_2^n(k_2|k_0))\notin\mathcal{T}_{\epsilon'}^{(n)}  \\
&& \text{ for all } (k_0,k_1,k_2,\ell_1,\ell_2)\}, \\
\mathcal{E}_1 &=& \{(U_1^n(M_1,L_1),Y_1^n)\notin\mathcal{T}_\epsilon^{(n)} \}, \\
\mathcal{E}_2 &=& \{(U_1^n(m_1,\ell_1),Y_1^n)\in\mathcal{T}_\epsilon^{(n)}\text{ for some } (m_1,\ell_1)\neq(M_1,L_1)\}.
\end{IEEEeqnarray*}
The average error probability for $M_1$ is upper bounded as 
\begin{IEEEeqnarray*}{rCl}
\mathbb{P}(\{\hat{M_1}\neq M_1\}) &\le& \mathbb{P}(\mathcal{E}_0) + \mathbb{P}(\mathcal{E}_1\cap\mathcal{E}_0^c) + \mathbb{P}(\mathcal{E}_2).
\end{IEEEeqnarray*}

By extending \cite[Lemma 14.1, p. 351]{ElGamal:11}, it can be shown that the term $\mathbb{P}(\mathcal{E}_0)$ tends to zero as $n\to\infty$ if $\tilde{R}_1+\tilde{R}_2 > I(U_1;U_2) + \delta(\epsilon')$ and 
\begin{IEEEeqnarray*}{rCl}
R'_0 + \sum_{k\in\mathcal{S}} R'_k + \sum_{j\in\mathcal{D}} \tilde{R}_j &>& 
I(U(\mathcal{D});X_0,X(\mathcal{S})) + \mathbbold{1}\{\mathcal{S}=\{1,2\}\}I(X_1;X_2|X_0) \\
&& + \mathbbold{1}\{\mathcal{D}=\{1,2\}\}I(U_1;U_2) + \delta(\epsilon'), 
\end{IEEEeqnarray*}
for all $\mathcal{D},\mathcal{S}\subseteq\{1,2\}$.
Next, due to the codebook construction and the conditional typicality lemma~\cite[p. 27]{ElGamal:11}, $\mathbb{P}(\mathcal{E}_1\cap\mathcal{E}_0^c)$ tends to zero as $n\to\infty$. Finally, using the joint typicality lemma~\cite[p. 29]{ElGamal:11}, $\mathbb{P}(\mathcal{E}_2)$ tends to zero as $n\to\infty$ if 
\begin{IEEEeqnarray*}{rCl}
R_1 + \tilde{R}_1  &<& I(U_1;Y_1) - \delta(\epsilon).
\end{IEEEeqnarray*}

The average error probability for $M_2$ can be bounded in a similar manner and then we have the additional rate condition
\begin{IEEEeqnarray*}{rCl}
R_2 + \tilde{R}_2  &<& I(U_2;Y_2) - \delta(\epsilon).
\end{IEEEeqnarray*}
Using the Fourier--Motzkin elimination to project out $\tilde{R}_1$, $\tilde{R}_2$, and $R'_{0j}, R'_j$, $j\in\{0,1,2\}$, we obtain the rate conditions in Theorem \ref{thm:compress}. Finally, the theorem is established by letting $\epsilon\to 0$.
\end{IEEEproof}

\subsection{Distributed Decode--Forward for Broadcast} \label{subsec:DDF}
The DDF scheme for broadcast~\cite{Lim:15}, which is developed for general memoryless broadcast relay networks, in particular applies to downlink C-RAN with arbitrary $N$ BSs and $L$ users.\footnote{The problem statement in Section \ref{sec:prob_state} has to be expanded to general number of BSs and users and to allow symbol-wise operations.} The following theorem states its performance in this setup. For convenience, we denote $\tilde{X}=(W_1,\cdots,W_N)$, $\breve{X}_j=(X_j,(W_{kj}:k\neq j))$, $j\in[N]$, and $\breve{Y}_k=(W_k,(W_{kj}:j\neq k))$, $k\in[N]$. 

\begin{theorem}[Lim, Kim, Kim{\cite[Theorem 2]{Lim:15}}] \label{thm:DDF}
A rate tuple $(R_1,\cdots,R_L)$ is achievable for the downlink $N$-BS $L$-user C-RAN with BS cooperation if 
\begin{IEEEeqnarray*}{rCl} \label{eq:DDF_ori}
\sum_{\ell\in\mathcal{D}} R_\ell &<& I(\tilde{X},\breve{X}(\mathcal{S});\tilde{U}(\mathcal{S}^c),U(\mathcal{D})|\breve{X}(\mathcal{S}^c)) 
- \sum_{k\in\mathcal{S}^c}\left[ I(\tilde{U}_k;\tilde{U}(\mathcal{S}_k^c),\tilde{X},\breve{X}^N|\breve{X}_k,\breve{Y}_k)+I(\breve{X}_k;\breve{X}(\mathcal{S}_k^c))\right] \\
\label{eq:DDF}
&& -\sum_{\ell\in\mathcal{D}} I(U_\ell;U(\mathcal{D}_\ell),\tilde{U}(\mathcal{S}^c),\tilde{X},\breve{X}^N|Y_\ell), \IEEEyesnumber
\end{IEEEeqnarray*}
for all $\mathcal{S}\subseteq[N]$, $\mathcal{D}\subseteq[L]$ for some pmf $p_{\tilde{U}^N,U^L,\tilde{X},\breve{X}^N}$, where $\mathcal{S}_k^c=\mathcal{S}^c\cap[k-1]$ and $\mathcal{D}_\ell=\mathcal{D}\cap[\ell-1]$.
\end{theorem}

The following proposition shows that, in Theorem~\ref{thm:DDF}, it is without loss in optimality to restrict the distribution of $(\tilde{U}^N,U^L,\tilde{X},\breve{X}^N)$ in the following manner:  
\begin{enumerate}
\item $p_{\tilde{U}^N,U^L,\tilde{X},(X^N,\{W_{kj}\})}=p_{\tilde{U}^N,\tilde{X},\{W_{kj}\}}p_{U^L,X^N}$;
\item $\tilde{U}_k=(W_k,(W_{kj}:j\neq k))$; 
\item $p_{W^N,\{W_{kj}\}}=\prod_{k=1}^Np_{W_k}\prod_{(j,k):j\neq k}p_{W_{kj}}$;
\item $W_k\sim$ Uniform($[2^{C_k}]$); and 
\item $W_{kj}\sim$ Uniform($[2^{C_{kj}}]$).
\end{enumerate}

\begin{proposition}\label{prop:DDF}
A rate tuple $(R_1,\cdots,R_L)$ lies in the achieved rate region~\eqref{eq:DDF_ori} of the DDF scheme for the downlink $N$-BS $L$-user C-RAN with BS cooperation if and only if there exists some joint pmf $p_{X^N,U^L}$ such that 
\begin{IEEEeqnarray}{rCl} \label{eq:DDF_refine}
\sum_{\ell\in\mathcal{D}} R_\ell &<& \sum_{\ell\in\mathcal{D}}I(U_\ell;Y_\ell) + \sum_{k\in\mathcal{S}^c} C_k + \sum_{j\in\mathcal{S}}\sum_{k\in\mathcal{S}^c} C_{kj} - \Gamma(X(\mathcal{S}^c),U(\mathcal{D})), \label{eq:DDF_refined}
\end{IEEEeqnarray}
for all $\mathcal{S}\subseteq[N]$ and $\mathcal{D}\subseteq[L]$ such that $|\mathcal{D}|\ge 1$. 
\end{proposition}

\begin{remark} For the case in which $N=2$ and $L=2$, it can be shown easily that setting $X_0=\emptyset$ in the rate region in Theorem~\ref{thm:compress} one recovers \eqref{eq:DDF_refine}. Thus, for the network model in Figure~\ref{fig:system_coop}, our generalized compression scheme (G-Compression) outperforms the DDF scheme for broadcast~\cite[Theorem 2]{Lim:15}.
\hfill$\lozenge$\end{remark}

Since downlink C-RAN is a special instance of memoryless broadcast relay networks, the DDF scheme achieves any point in the capacity region of an $N$-BS $L$-user C-RAN to within a gap of $(1+N+L)/2$ bits per dimension under the memoryless Gaussian model~\cite[Corollary 8]{Lim:15}. The following theorem tightens this gap for downlink C-RANs. The proof is deferred to Appendix~\ref{appendix:gap}.

\begin{theorem} \label{thm:gap}
Consider the downlink of any $N$-BS $L$-user C-RAN with BS cooperation.
Under the memoryless Gaussian model, the DDF scheme for broadcast achieves within $\frac{L}{2}+\frac{\min\{N,L\log N\}}{2}$ bits per dimension from the capacity region.
\end{theorem}

\section{Comparison and Numerical Evaluations} \label{sec:compare_evaluate}

In this section, we evaluate and compare our G-DS scheme and G-Compression scheme in two useful examples. In the first example, the G-DS scheme (as well as the data-sharing scheme) is optimal, whereas the G-Compression scheme is strictly suboptimal. In the second example the opposite is true. In the second part of this section, we provide numerical results for the memoryless Gaussian model.

\subsection{Examples} \label{subsec:ex_DSvsC}
\begin{example}[One BS and One User] \label{ex:1bs1mu}
Consider the special case with only one BS and one user, as depicted in Figure~\ref{fig:1b1u}. (Our model reduces to this scenario when the DM-IC is of the form $p_{Y_1,Y_2|X_1,X_2}=p_{Y_1,Y_2|X_1}$ and when $C_2=R_2=0$.)  Decode-and-forward~\cite{Cover:79} is optimal in this special case and rate $R_1$ is achievable whenever
\begin{IEEEeqnarray*}{rCl}
R_1 &<& \min\left\{C_1,\max_{p_{X_1}}I(X_1;Y_1)\right\}.
\end{IEEEeqnarray*}
Furthermore, compress-and-forward~\cite{Cover:79} is also optimal since the first hop is noiseless. 
This performance is also recovered by the G-DS scheme; see Corollary~\ref{col:1b2u} specialized to $R_2=0$ and the choice of auxiliaries  $V=\emptyset$ and $X_1=U$.  

\begin{figure}[t!]
\begin{center}
\includegraphics[scale=0.8, trim={0.5cm 0.5cm 7cm 27.2cm}, clip]{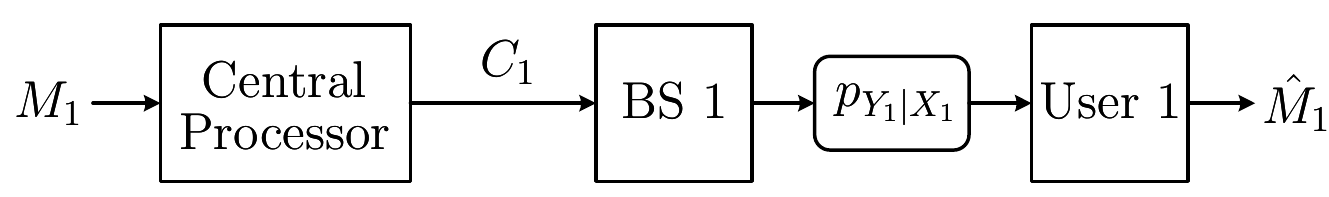}
\end{center}
\vspace{-0.1in}
\caption{The system considered in Example \ref{ex:1bs1mu}.}
\label{fig:1b1u}
\vspace{-0.1in}
\end{figure}

The G-Compression scheme and the DDF scheme for broadcast achieve all rates $R_1$ that satisfy:
\begin{IEEEeqnarray*}{rCl}
R_1&<& I(U_1;Y_1), \\
R_1 &<& C_1 + I(U_1;Y_1) - I(U_1;X_1) \\
&=& C_1 - I(U_1;X_1|Y_1), 
\end{IEEEeqnarray*}
for some pmf $p_{U_1,X_1}$ s.t. $U_1\markov X_1\markov Y_1$ form a Markov chain. 

If the second hop is deterministic, i.e., $Y_1$ is a deterministic function of $X_1$, then 
the G-Compression scheme with $U_1=Y_1$ achieves the capacity. 

However, if the second hop is not deterministic, then setting $U_1=Y_1$ violates the Markov condition $U_1\markov X_1\markov Y_1$. In general, the G-Compression scheme is  suboptimal. To see this, consider a DM-IC satisfying $p_{Y_1|X_1}(y_1|x_1) < 1$ for all $(x_1,y_1)\in\mathcal{X}_1\times\mathcal{Y}_1$, i.e., for all inputs $x_1\in\mathcal{X}_1$, the output~$Y_1$ is not a deterministic function of $x_1$. Then, for every pmf $p_{X_1}$, the corresponding joint pmf $p_{X_1,Y_1}$ is indecomposable.\footnote{A joint pmf $p_{X,Y}$ is said to be {\em indecomposable} \cite[Problem 15.12, p. 345]{Csiszar:11} if there are no functions $f$ and $g$ with respective domains $\mathcal{X}$ and $\mathcal{Y}$ so that 1) $\mathbb{P}(f(X)=g(Y))=1$ and 2) $f(X)$ takes at least two values with non-zero probability.} Next, let us assume that $0 < C_1 < \max_{p_{X_1}}I(X_1;Y_1)$. Now we show that the G-Compression scheme is not capacity achieving by contradiction. 

If the G-Compression scheme is capacity achieving, then it holds that the capacity-achieving distribution $p_{U_1,X_1}$ satisfies that $I(U_1;X_1|Y_1)=0$, i.e., $U_1\markov Y_1 \markov X_1$ form a Markov chain. However, since  $U_1\markov X_1 \markov Y_1$ also form a Markov chain, the indecomposability of the joint pmf $p_{X_1,Y_1}$ implies that the capacity-achieving distribution $p_{U_1,X_1}$ satisfies that $U_1$ is independent of $(X_1,Y_1)$ (see \cite[Problem 16.25, p. 392]{Csiszar:11}) and thus $I(U_1;Y_1)=0$, which contradicts that the joint pmf $p_{U_1,X_1}$ achieves the capacity $C_1>0$.
\hfill$\lozenge$\end{example}

\begin{figure}[t!]
\begin{center}
\includegraphics[scale=0.8, trim={0.5cm 0.5cm 6cm 24.2cm}, clip]{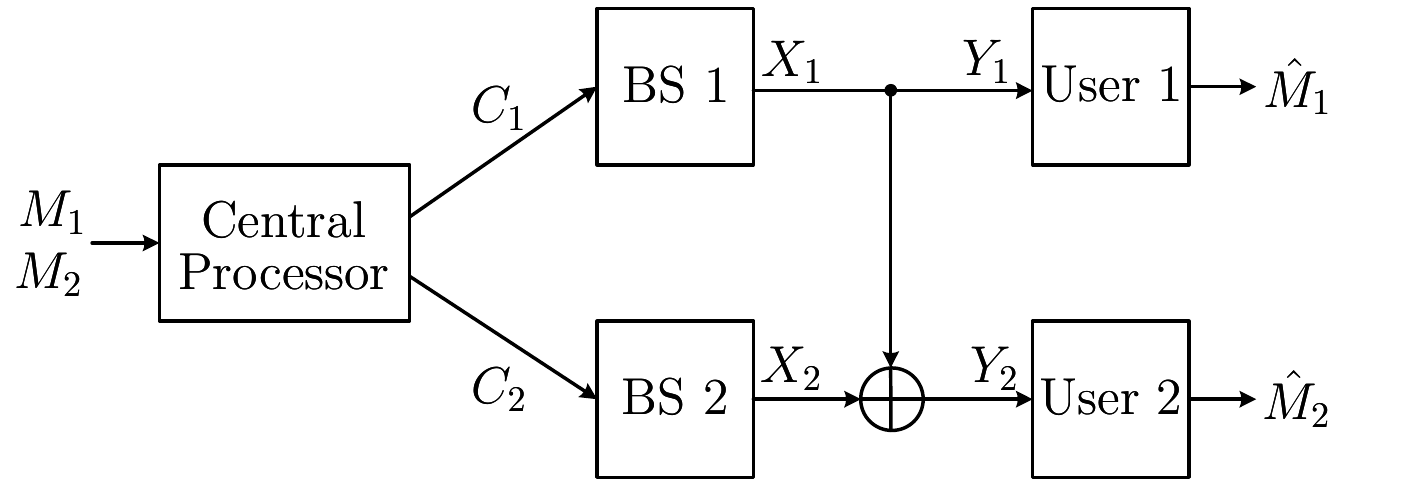}
\end{center}
\vspace{-0.15in}
\caption{The system considered in Example \ref{ex:Zchannel}.}
\label{fig:Zch}
\vspace{-0.1in}
\end{figure}

\begin{example}[Z-Interference Channel] \label{ex:Zchannel}
Consider the case where $C_1=C_2=1$, $C_{12}=C_{21}=0$, $\mathcal{X}_1=\mathcal{X}_2=\{0,1\}$, $Y_1=X_1$, and $Y_2=X_1\oplus X_2$. The system is depicted in Figure~\ref{fig:Zch}. Now we show that the rate pair $(R_1,R_2)=(1,1)$, which is on the boundary of the capacity region, is achievable by the G-Compression scheme but not by the G-DS scheme.

 The following scheme achieves the desired rate pair $(R_1, R_2)=(1,1)$. Fix a blocklength $n$ and denote by $B_{\ell}^n:=(B_{\ell,1}, \ldots, B_{\ell,n})$ the $n$-bits  representation of $M_\ell$, $\ell\in\{1,2\}$.  The central processor sends all bits $B_1^n$ to BS~1, and it sends the x-or bits $B_{\oplus}^n:=(B_{1,1}\oplus B_{2,1}, \ldots, B_{1,n}\oplus B_{2,n})$ to BS~2. BS~1 sends inputs $X_1^n=B_1^n$ over the DM-IC and BS~2 sends inputs $X_2^n=B_{\oplus}^n$. 
 
 The same performance is achieved by the G-Compression scheme when the auxiliaries $(U_1,U_2)$ are chosen i.i.d. Bernoulli($1/2$), and $X_0=\emptyset$, $X_1=U_1$, and $X_2=U_1\oplus U_2$.

Now let us investigate the G-DS scheme. We consider the following relaxed conditions, where the inequalities do not need to be strict:
\begin{IEEEeqnarray*}{rCl}
R_1+R_2 &\overset{(a)}{\le}& I(U_0,U_1,U_2;Y_1) + I(V_0,V_1,V_2;Y_2) - I(U_0,U_1,U_2;V_0,V_1,V_2), \\
2R_1+R_2 &\overset{(b)}{\le}& C_1+C_2+C_{12}+C_{21} + I(U_1,U_2;Y_1|U_0) - I(U_1,V_1;U_2,V_2|U_0,V_0), 
\end{IEEEeqnarray*}
where $(a)$ follows by combining \eqref{eq:Dcond1}, \eqref{eq:Dcond2}, and \eqref{eq:Dcond3} with $\Omega_{\sf u}=\Omega_{\sf v}=\{0,1,2\}$ and $(b)$ follows by combining two times of \eqref{eq:Dcond1} with $(\Omega_{\sf u},\Omega_{\sf v})=(\{0,1,2\},\{0,1,2\})$ and $(\Omega_{\sf u},\Omega_{\sf v})=(\{0,1,2\},\emptyset)$, \eqref{eq:Dcond2} with $\Omega_{\sf u}=\{1,2\}$, \eqref{eq:Dcond4}, and \eqref{eq:Dcond5}. 

If $(R_1,R_2)=(1,1)$ is achievable by the G-DS scheme, then there must exist a joint pmf $p_{U_0,V_0,U_1,V_1,U_2,V_2}$ and functions $x_k(u_0,v_0,u_k,v_k)$, $k\in\{1,2\}$, such that 
\begin{enumerate}[leftmargin=*]
\item $I(U_0,U_1,U_2;V_0,V_1,V_2) = 0$;
\item $I(U_1,V_1;U_2,V_2|U_0,V_0) = 0$; 
\item $I(U_0,U_1,U_2;Y_1) = 1$;
\item $I(V_0,V_1,V_2;Y_2) = 1$; 
\item $I(U_1,U_2;Y_1|U_0) = 1$.
\end{enumerate}
However, the above constraints cannot be satisfied simultaneously. To see this, let us assume that the first four conditions hold, which imply that
\begin{enumerate}[leftmargin=*]
\item $(U_0,U_1,U_2)$ is independent of $(V_0,V_1,V_2)$;
\item the Markov chains $U_1\markov U_0\markov U_2$ and $V_1\markov V_0\markov V_2$ hold; and
\item $H(X_1|U_0,U_1,U_2)=H(X_1\oplus X_2|V_0,V_1,V_2)=0$.
\end{enumerate}
Thus, 
\begin{IEEEeqnarray*}{rCl}
I(X_1;V_0,V_1|U_0,U_1) &=& I(X_1,U_0,U_1;V_0,V_1) \\
&\le& I(X_1,U_0,U_1,U_2;V_0,V_1), \\
&\overset{(a)}{=}& I(U_0,U_1,U_2;V_0,V_1) = 0, 
\end{IEEEeqnarray*}
where $(a)$ follows since $H(X_1|U_0,U_1,U_2)=0$.
Since $X_1$ is a function of $(U_0,V_0,U_1,V_1)$ by construction, we have 
$H(X_1|U_0,U_1)=H(X_1|U_0,V_0,U_1,V_1)=0$, i.e., $X_1$ is a function of $(U_0,U_1)$. Finally, it holds that 
\begin{IEEEeqnarray*}{rCl}
0 &=& H(X_1\oplus X_2|V_0,V_1,V_2) \\
&\ge& H(X_1\oplus X_2|U_0,U_2,V_0,V_1,V_2) \\
&=& H(X_1|U_0,U_2,V_0,V_1,V_2) \\
&\overset{(a)}{=}& H(X_1|U_0), 
\end{IEEEeqnarray*}
where $(a)$ follows since $X_1$ is a function of $(U_0,U_1)$; $(U_0,U_1,U_2)$ is independent of $(V_0,V_1,V_2)$; and $U_1\markov U_0\markov U_2$ form a Markov chain. From all above we obtain that constraint $5)$ cannot be satisfied since $Y_1$ is a function of $U_0$, which concludes that the G-DS scheme cannot achieve the rate pair $(1,1)$.
\hfill$\lozenge$\end{example}

\subsection{Numerical Evaluation for the Memoryless Gaussian Model}\label{subsec:Gaussian}
In this subsection, we compare the achieved sum rates of the various coding schemes under the memoryless Gaussian model. We are mainly interested in the scenarios where the G-DS scheme outperforms the G-Compression scheme and the reverse compute--forward. Evaluating the G-DS scheme directly is challenging, so we evaluate the special cases with restricted correlation structures and then apply time sharing on them. To summarize, we evaluate the following schemes
\begin{enumerate}
\item G-DS scheme I, II, and III (Corollaries \ref{col:dsI}, \ref{col:dsII}, and \ref{col:dsIII}),
\item G-Compression scheme (Theorem \ref{thm:compress}), and 
\item reverse compute--forward with power allocation\cite{Hong:13}.
\end{enumerate}
For simplicity, we consider the symmetric case, i.e., $C_1=C_2=C$, $C_{12}=C_{21}=T$, $g_{11}=g_{22}=1$, and $|g_{12}|=|g_{21}|$. Then, the achievable sum rate $R_1+R_2$ can be upper bounded using the cut-set bound as  
\begin{IEEEeqnarray}{rCl} \label{eq:cut_ref}
R_1+R_2 &<& \min\{2C,R_{\sf sum}^\star\},
\end{IEEEeqnarray}
where $R_{\sf sum}^\star$ denotes the optimal sum rate assuming $C=\infty$, which can be computed by evaluating the corresponding Gaussian MIMO broadcast channel. We will use the cut-set bound~\eqref{eq:cut_ref} as a reference for comparison.

Now let us specify our choice of auxiliary random variables for the various schemes. Except for the G-DS scheme II, all other schemes are evaluated based on dirty paper coding. Let $\mathbf{S}^{(k)}$ be a $2\times 1$ jointly Gaussian random vector with zero-mean entries and covariance matrix $\mathsf{K}^{(k)}$, for $k\in\{1,2\}$. We assume that $\mathbf{S}^{(1)}$ and $\mathbf{S}^{(2)}$ are independent. For notational convenience, we denote $\mathbf{g}_2 = \begin{bmatrix} g_{21} & g_{22} \end{bmatrix}$.
\begin{enumerate}
\item Description I: $U_0=\mathbf{S}^{(1)}$, $V_0=\mathbf{S}^{(2)} + \mathsf{A}\mathbf{S}^{(1)}$, and $\begin{bmatrix} X_1 \\ X_2 \end{bmatrix} = \mathbf{S}^{(1)} + \mathbf{S}^{(2)}$, where 
\begin{IEEEeqnarray*}{rCl} 
\mathsf{A} &=& \mathsf{K}^{(2)}\mathbf{g}_2^T\left(1+\mathbf{g}_2\mathsf{K}^{(2)}\mathbf{g}_2^T\right)^{-1}\mathbf{g}_2.
\end{IEEEeqnarray*}
Note that $X_k=U_k+V_k$, $k\in\{1,2\}$. We optimize over the covariance matrices $\mathsf{K}^{(1)}$ and $\mathsf{K}^{(2)}$ that satisfy the average power constraints. 
\item Description II: The random variables $(U_0,V_0,U_1,V_1,U_2,V_2)$ are i.i.d. $\mathcal{N}(0,1)$ and $X_1=a_1U_0+a_2V_0+a_3U_1+a_4V_1$ and $X_2=b_1U_0+b_2V_0+b_3U_2+b_4V_2$ for some $a_j,b_j\in\mathbb{R}$, $j\in\{1,2,3,4\}$. We optimize over the coefficients $(a_j,b_j:j\in\{1,2,3,4\})$ that satisfy the average power constraints.
\item Description III: 
\begin{IEEEeqnarray*}{rCl}
\begin{bmatrix} U_1 \\ U_2 \end{bmatrix} &=& \mathbf{S}^{(1)}, \\
\begin{bmatrix} V_1 \\ V_2 \end{bmatrix} &=& \mathbf{S}^{(2)} + \mathsf{A}\mathbf{S}^{(1)}, \\
\begin{bmatrix} X_1 \\ X_2 \end{bmatrix} &=& (\mathsf{I}+\mathsf{A})\mathbf{S}^{(1)} + \mathbf{S}^{(2)}, 
\end{IEEEeqnarray*}
where $\mathsf{I}$ is the $2\times2$ identity matrix and $\mathsf{A}$ is a $2\times2$ real-valued matrix. Note that $X_k=U_k+V_k$, $k\in\{1,2\}$.\footnote{We remark that since the BSs do not have full information about $\mathbf{S}^{(1)}$ and $\mathbf{S}^{(2)}$, setting $\begin{bmatrix} X_1 \\ X_2 \end{bmatrix} = \mathbf{S}^{(1)} + \mathbf{S}^{(2)}$ is not allowed because the resulting $X_k$ is not a function of $(U_k,V_k)$, $k\in\{1,2\}$. Also, due to this fact, the precoding matrix $\mathsf{A} = \mathsf{K}^{(2)}\mathbf{g}_2^T\left(1+\mathbf{g}_2\mathsf{K}^{(2)}\mathbf{g}_2^T\right)^{-1}\mathbf{g}_2$ is in general suboptimal.} We optimize over the covariance matrices $\mathsf{K}^{(1)}$, $\mathsf{K}^{(2)}$ and the precoding matrix $\mathsf{A}$ that satisfy the average power constraints.
\item Compression: $U_1=\mathbf{S}^{(1)}$, $U_2=\mathbf{S}^{(2)} + \mathsf{A}\mathbf{S}^{(1)}$, and $\begin{bmatrix} X_1 \\ X_2 \end{bmatrix} = \mathbf{S}^{(1)} + \mathbf{S}^{(2)} + \mathbf{W}$, where 
\begin{IEEEeqnarray*}{rCl} 
\mathsf{A} &=& \mathsf{K}^{(2)}\mathbf{g}_2^T\left(1+\mathbf{g}_2(\mathsf{K}^{(2)}+\mathsf{K}^{({\sf w})})\mathbf{g}_2^T\right)^{-1}\mathbf{g}_2, 
\end{IEEEeqnarray*}
and $\mathbf{W}$ is a $2\times 1$ jointly Gaussian random vector with zero-mean entries and covariance matrix $\mathsf{K}^{(\sf w)}$, independent of $(\mathbf{S}^{(1)},\mathbf{S}^{(2)})$. Finally, we let $X_0$ be an $\mathcal{N}(0,1)$ random variable such that $X_0$ and $(\mathbf{S}^{(1)},\mathbf{S}^{(2)},\mathbf{W})$ are jointly Gaussian. We optimize over the covariance matrices $\mathsf{K}^{(1)}$, $\mathsf{K}^{(2)}$, and $\mathsf{K}^{(\sf w)}$ that satisfy the average power constraints and over the covariances of $X_0$ with each of $(\mathbf{S}^{(1)},\mathbf{S}^{(2)},\mathbf{W})$.
\end{enumerate}

\begin{figure}[t!] 
\centering
\subfigure[$P=1$, $(g_{12},g_{21})=(0.5,0.5)$.]{\label{fig:Donly_P1_p05}\includegraphics[scale=0.55]{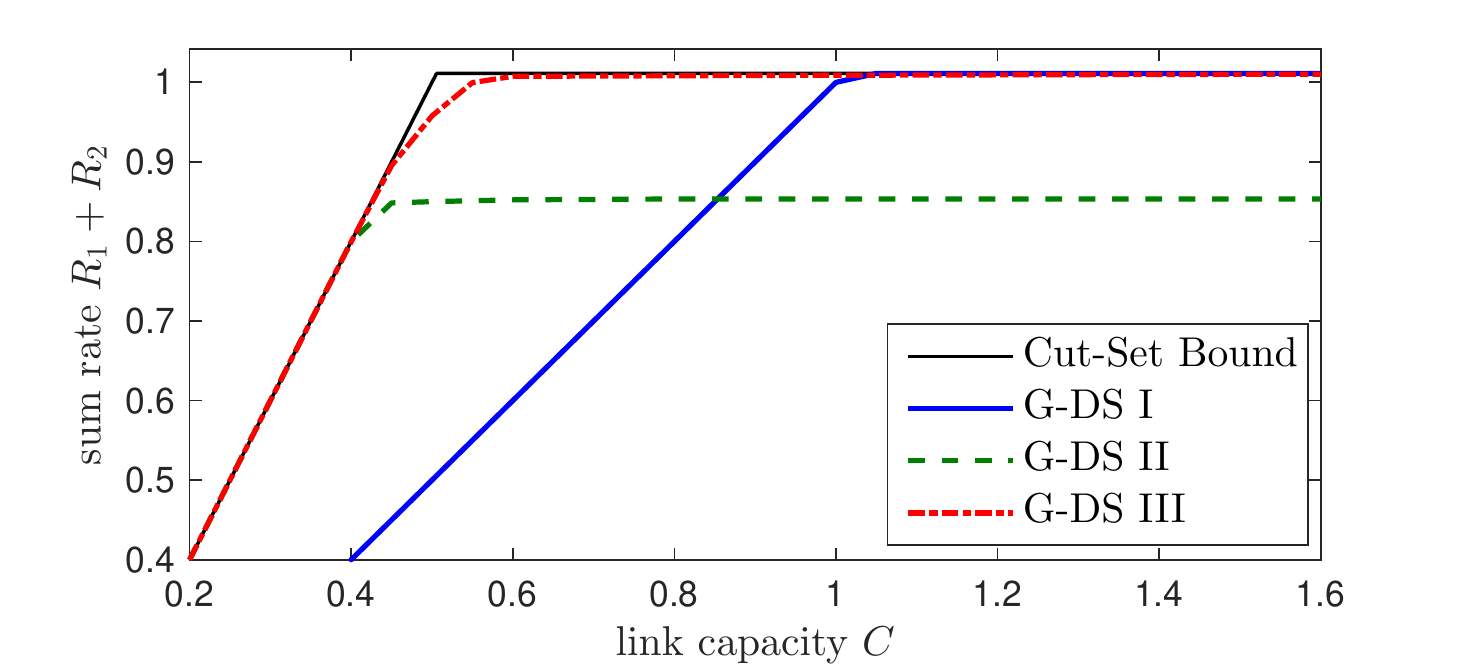}}
\subfigure[$P=1$, $(g_{12},g_{21})=(0.5,-0.5)$.]{\label{fig:Donly_P1_m05}\includegraphics[scale=0.55]{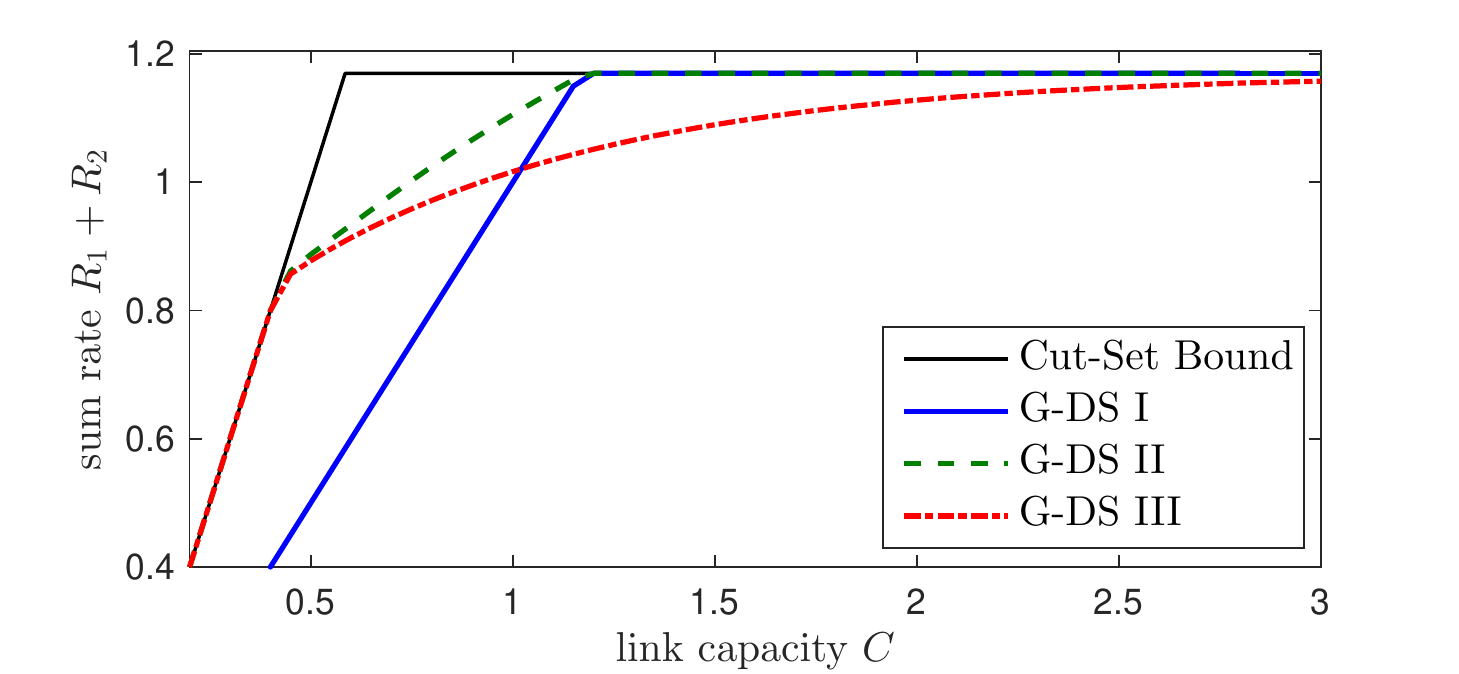}}

\subfigure[$P=100$, $(g_{12},g_{21})=(0.5,0.5)$.]{\label{fig:Donly_P100_p05}\includegraphics[scale=0.55]{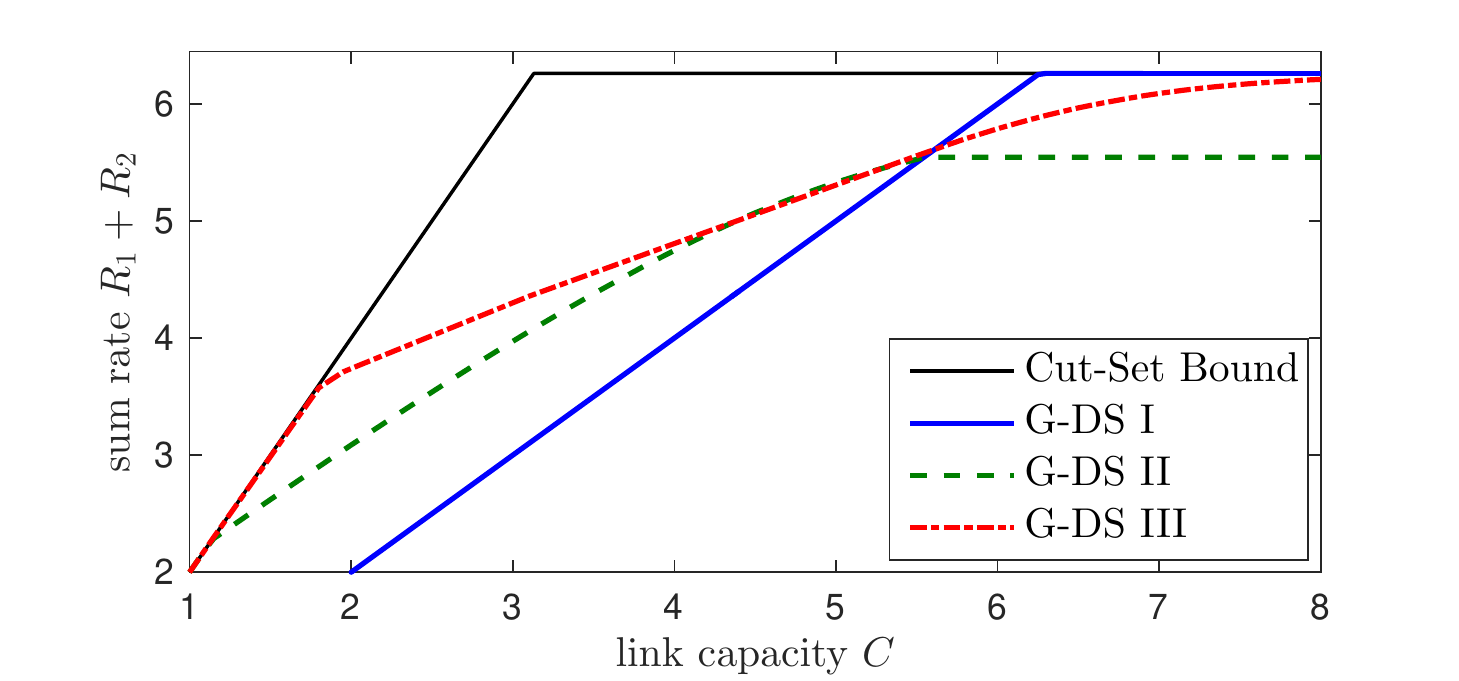}}
\subfigure[$P=100$, $(g_{12},g_{21})=(0.5,-0.5)$.]{\label{fig:Donly_P100_m05}\includegraphics[scale=0.55]{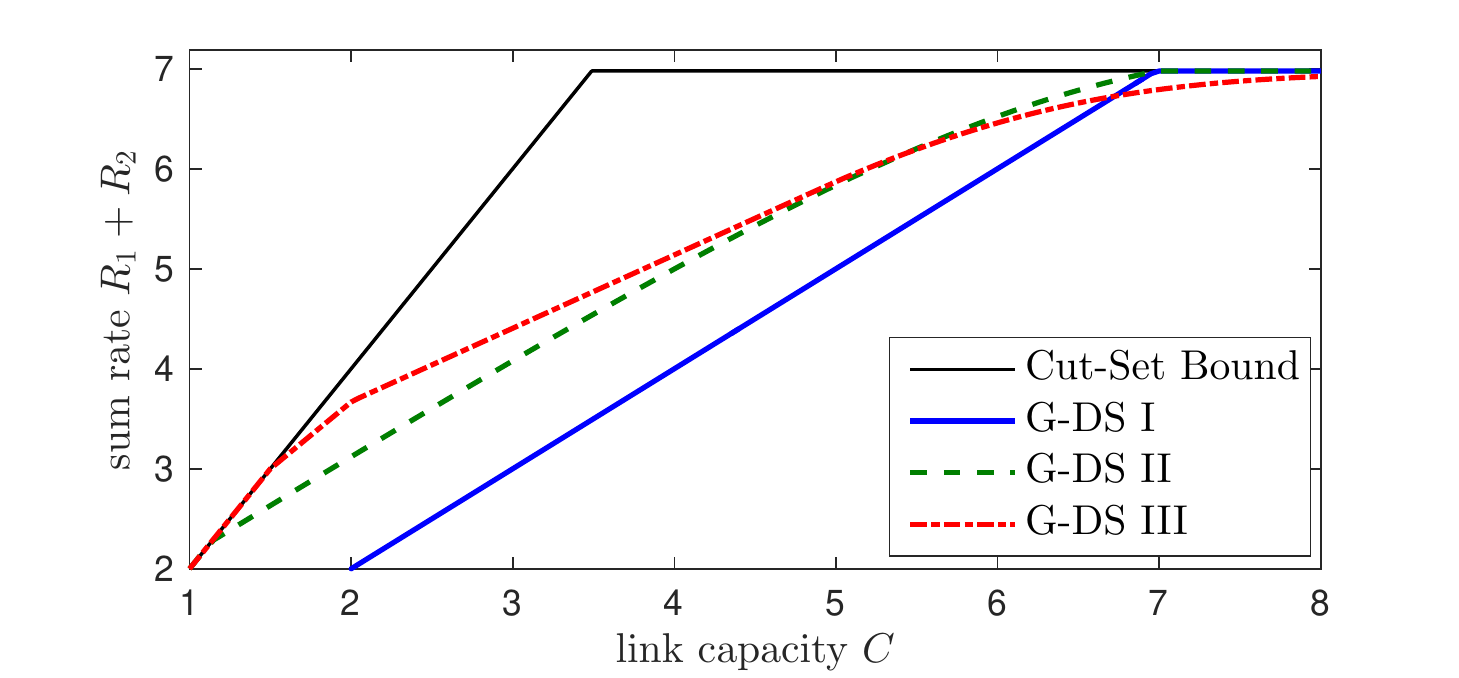}}
\caption{Achieved sum-rates of the G-DS schemes I, II, and III under the symmetric memoryless Gaussian model. Here $T=0$ and $g_{11}=g_{22}=1$.}
\label{fig:Donly}
\end{figure}

First, let us compare the G-DS schemes with different correlation structures. We assume that $T=0$. In Figure \ref{fig:Donly}, we fix $g_{12}=0.5$ and consider $(P,g_{21})\in\{1,100\}\times\{0.5,-0.5\}$. From the evaluation results, we make the following observations and remarks for the considered setup:  
\begin{itemize}[leftmargin=*]
\item In general, the G-DS scheme~I using only common codewords performs well in the strong-fronthaul regime, i.e., when $C$ is large. By contrast, the G-DS scheme~III using only private codewords performs well in the weak-fronthaul regime. \item Introducing correlation among codewords is useful. In fact, time sharing between the G-DS schemes~I and~III outperforms the G-DS scheme II for all values of link capacity $C$.
\item The G-DS scheme~III is more beneficial in the low-power regime, i.e., when $P$ is small.
\item When the channel gain matrix $G$ is well-conditioned, linear beamforming performs as good as dirty paper coding, which is the reason why the G-DS scheme~II outperforms the G-DS scheme~I in Figures~\ref{fig:Donly_P1_m05} and~\ref{fig:Donly_P100_m05}.
\end{itemize}
We remark that the achieved sum rate of the G-DS scheme~I can be simply expressed as $\min\{C+T,2C,R_{\sf sum}^\star\}$. Thus, when $T=0$, the G-DS scheme~I is optimal for the regime where $C\ge R_{\sf sum}^\star$.

\begin{figure}[t!] 
\centering
\subfigure[$P=1$, $(g_{12},g_{21})=(0.5,0.5)$.]{\label{fig:DCR_P1_p05}\includegraphics[scale=0.55]{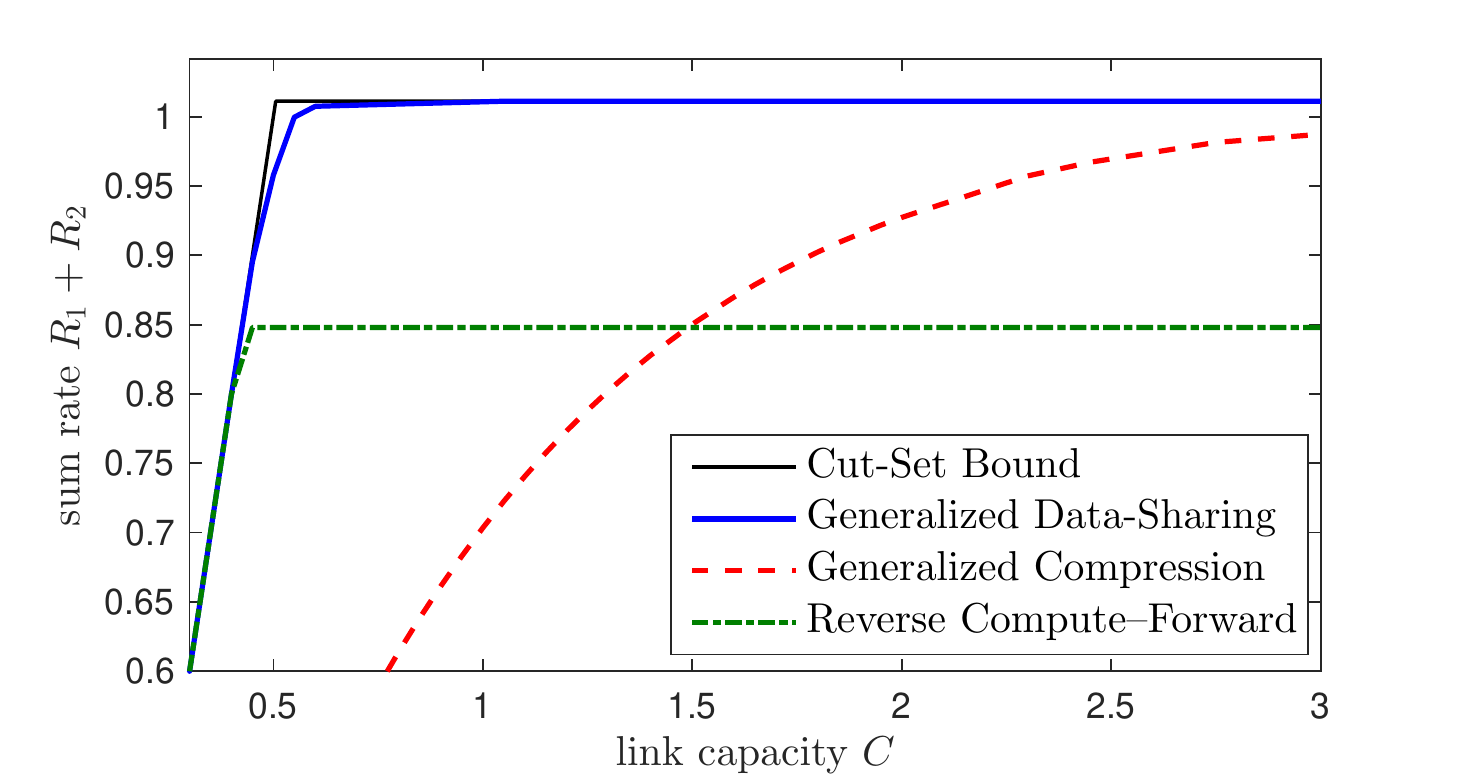}}
\subfigure[$P=1$, $(g_{12},g_{21})=(0.5,-0.5)$.]{\label{fig:DCR_P1_m05}\includegraphics[scale=0.55]{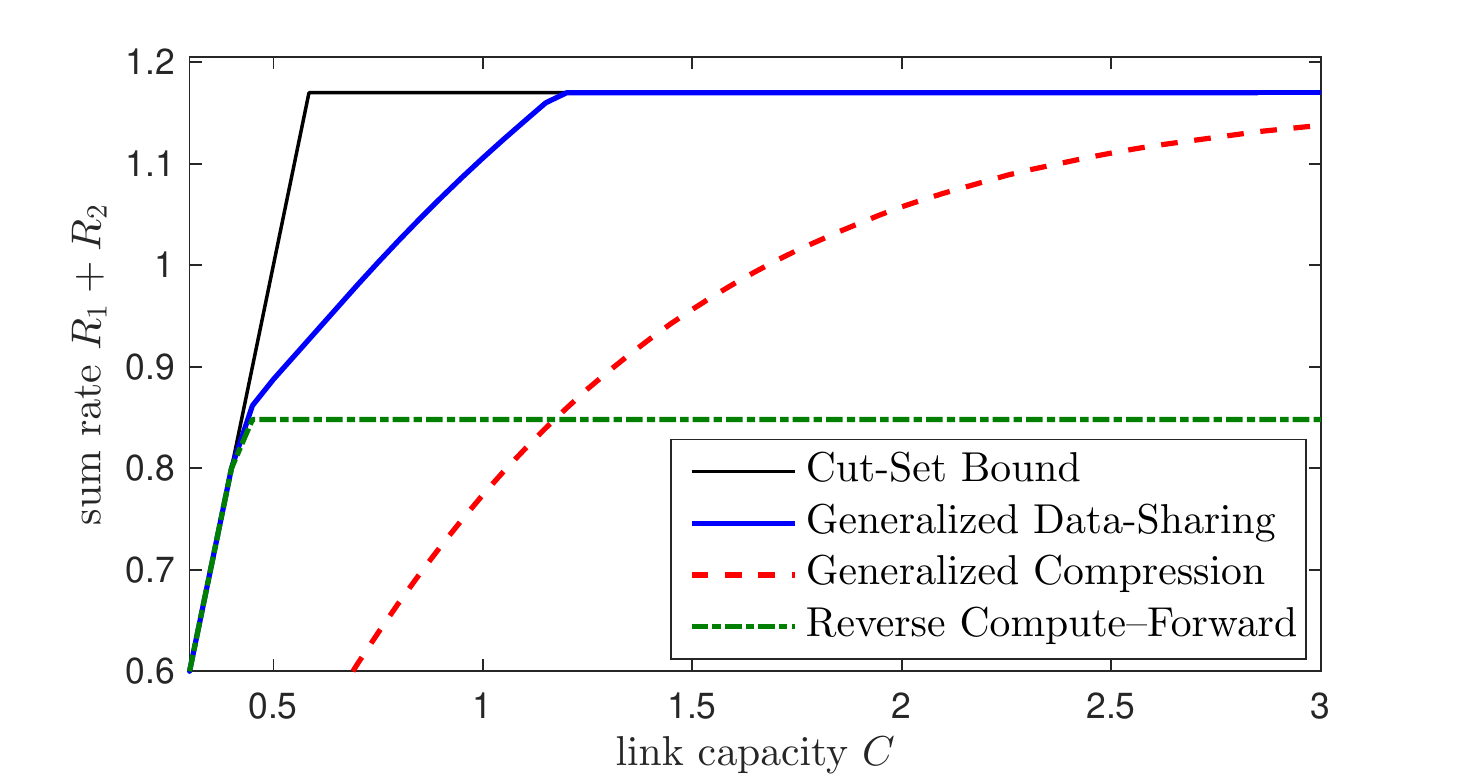}}

\subfigure[$P=10$, $(g_{12},g_{21})=(0.5,0.5)$.]{\label{fig:DCR_P10_p05}\includegraphics[scale=0.55]{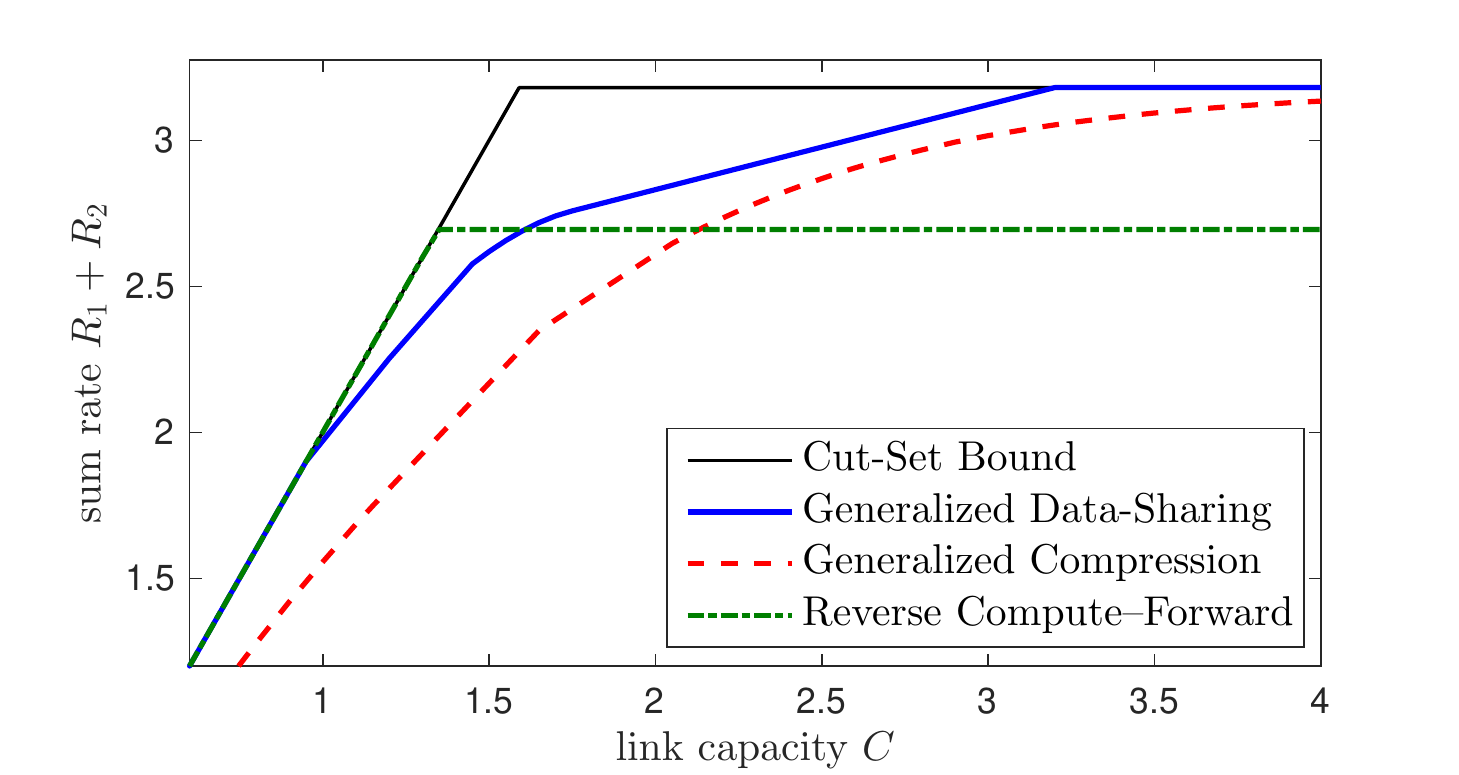}}
\subfigure[$P=10$, $(g_{12},g_{21})=(0.5,-0.5)$.]{\label{fig:DCR_P10_m05}\includegraphics[scale=0.55]{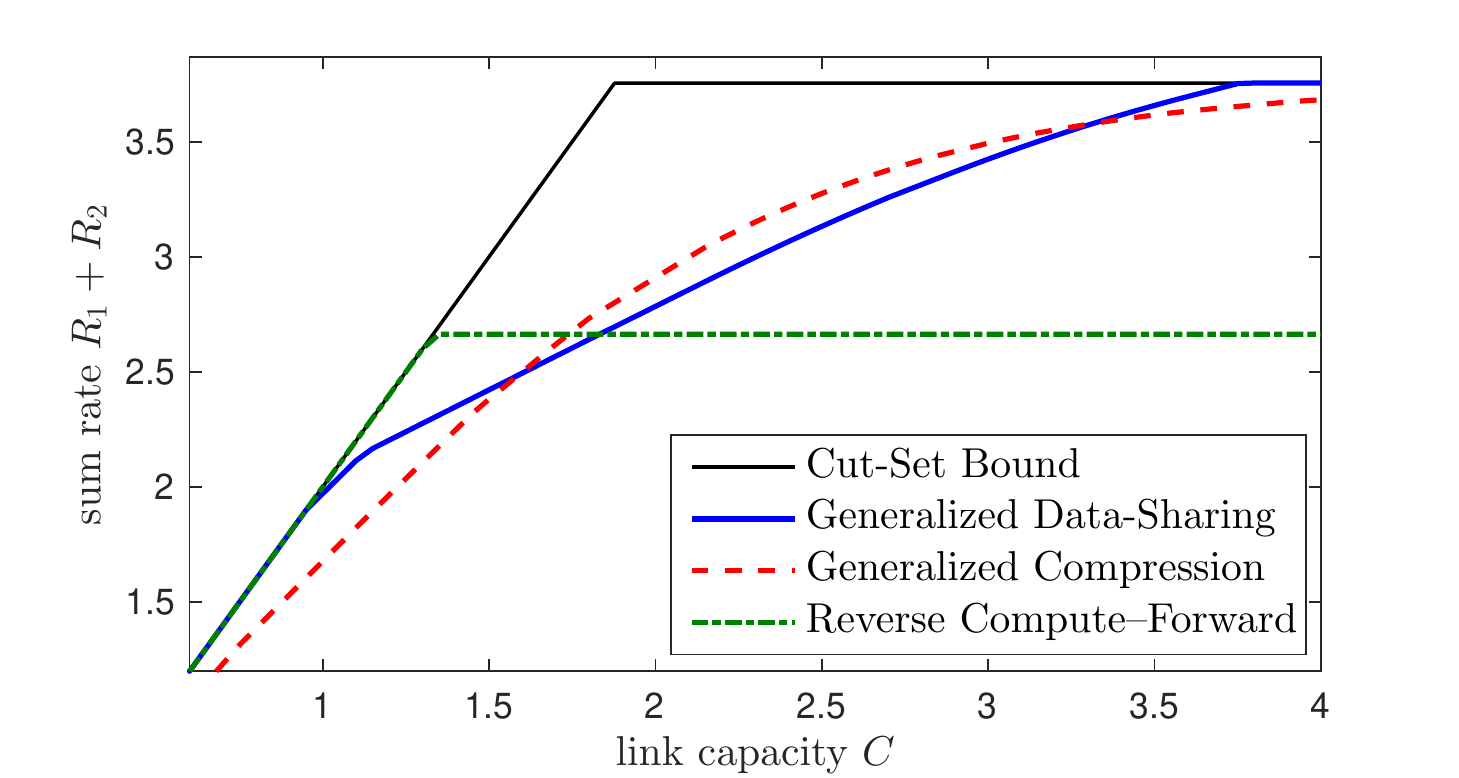}}

\subfigure[$P=100$, $(g_{12},g_{21})=(0.5,0.5)$.]{\label{fig:DCR_P100_p05}\includegraphics[scale=0.55]{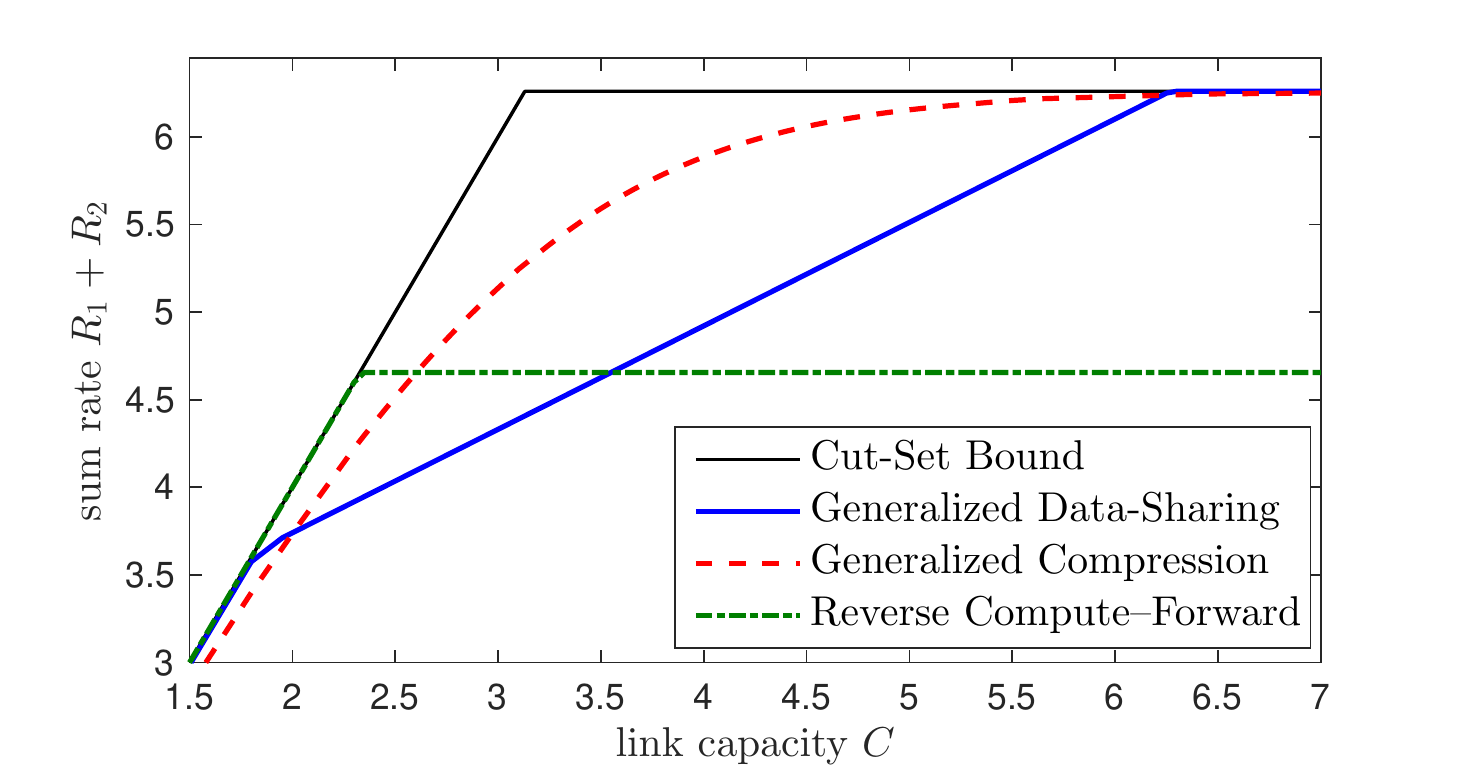}}
\subfigure[$P=100$, $(g_{12},g_{21})=(0.5,-0.5)$.]{\label{fig:DCR_P100_m05}\includegraphics[scale=0.55]{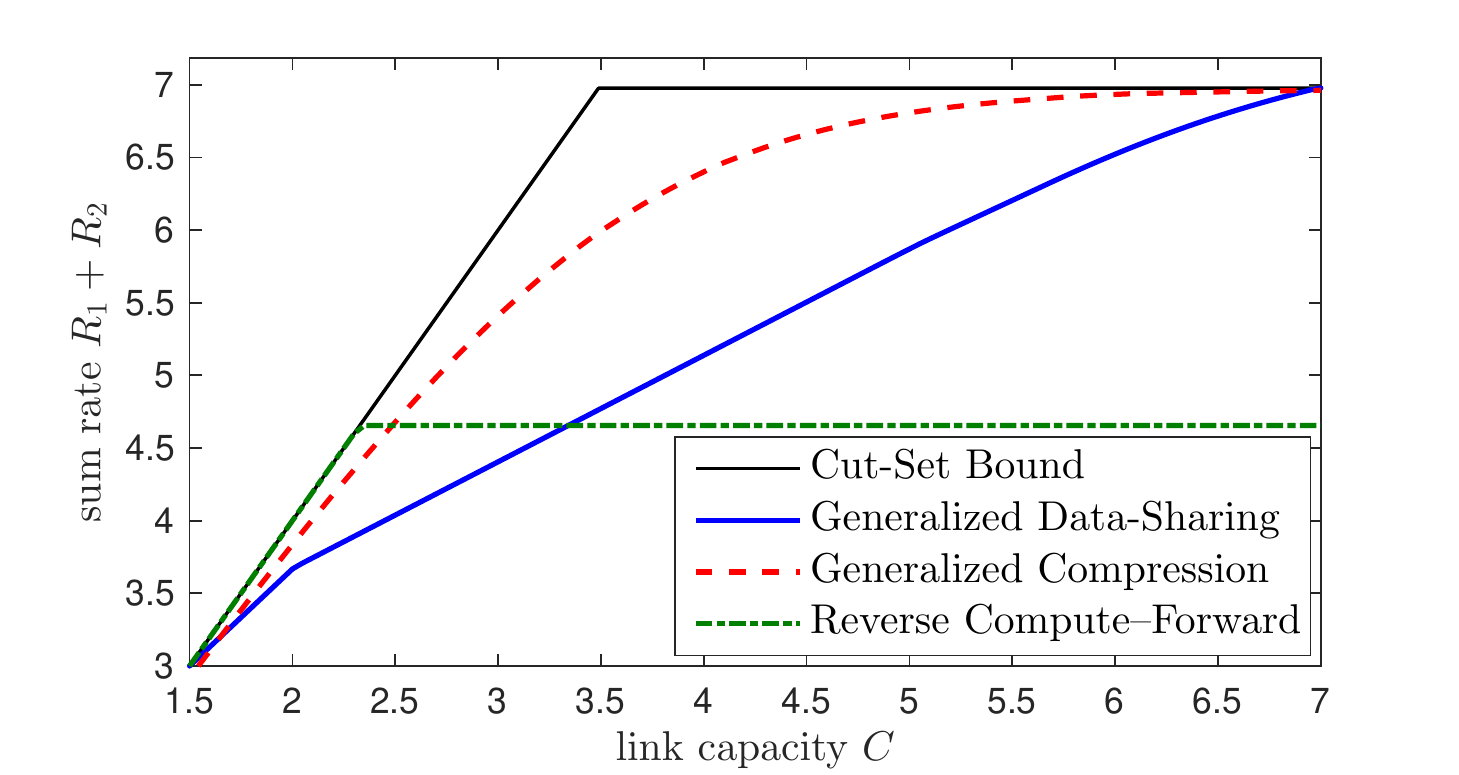}}

\caption{Achieved sum-rates of the G-DS scheme, the G-Compression scheme, and the reverse compute--forward scheme with power control under the symmetric memoryless Gaussian model. Here $T=0$ and $g_{11}=g_{22}=1$.}
\label{fig:DCR}
\end{figure}

Next, we compare the G-DS scheme (time sharing among the G-DS schemes I, II, and III) with the G-Compression scheme and the reverse compute--forward scheme. In Figure \ref{fig:DCR}, we fix $g_{12}=0.5$ and consider $(P,g_{21})\in\{1,10,100\}\times\{0.5,-0.5\}$. From the evaluation results, we make the following observations and remarks for the considered setup:  
\begin{itemize}[leftmargin=*]
\item The G-DS scheme achieves the optimal sum rate when the link capacity $C$ is relatively small or relatively large. The range of optimality depends on the power and the channel conditions. In general, in the low-power regime and/or when the channel gain matrix is ill-conditioned, the G-DS scheme has a more apparent advantage over the other two schemes.
\item The G-Compression scheme achieves a better performance in the high-power regime. As $P$ increases, the G-Compression scheme outperforms the other two schemes in the middle range of link capacity. 
\item The reverse compute--forward has a good performance when the link capacity $C$ is relatively small, especially when $P$ is large. However, the reverse compute--forward suffers from non-integer penalty and thus its achieved sum rate cannot reach $R_{\sf sum}^\star$ even if the link capacity~$C$ is large.
\end{itemize}

\newsavebox{\chmatsym}
\savebox{\chmatsym}{$\left[\begin{smallmatrix}g_{11}  & g_{12} \\ g_{21} & g_{22}\end{smallmatrix}\right]=\left[\begin{smallmatrix}1&0.5\\-0.5&1\end{smallmatrix}\right]$}

\begin{figure}[t!]
\vspace{-0.1in}
\begin{center}
\includegraphics[scale=0.8]{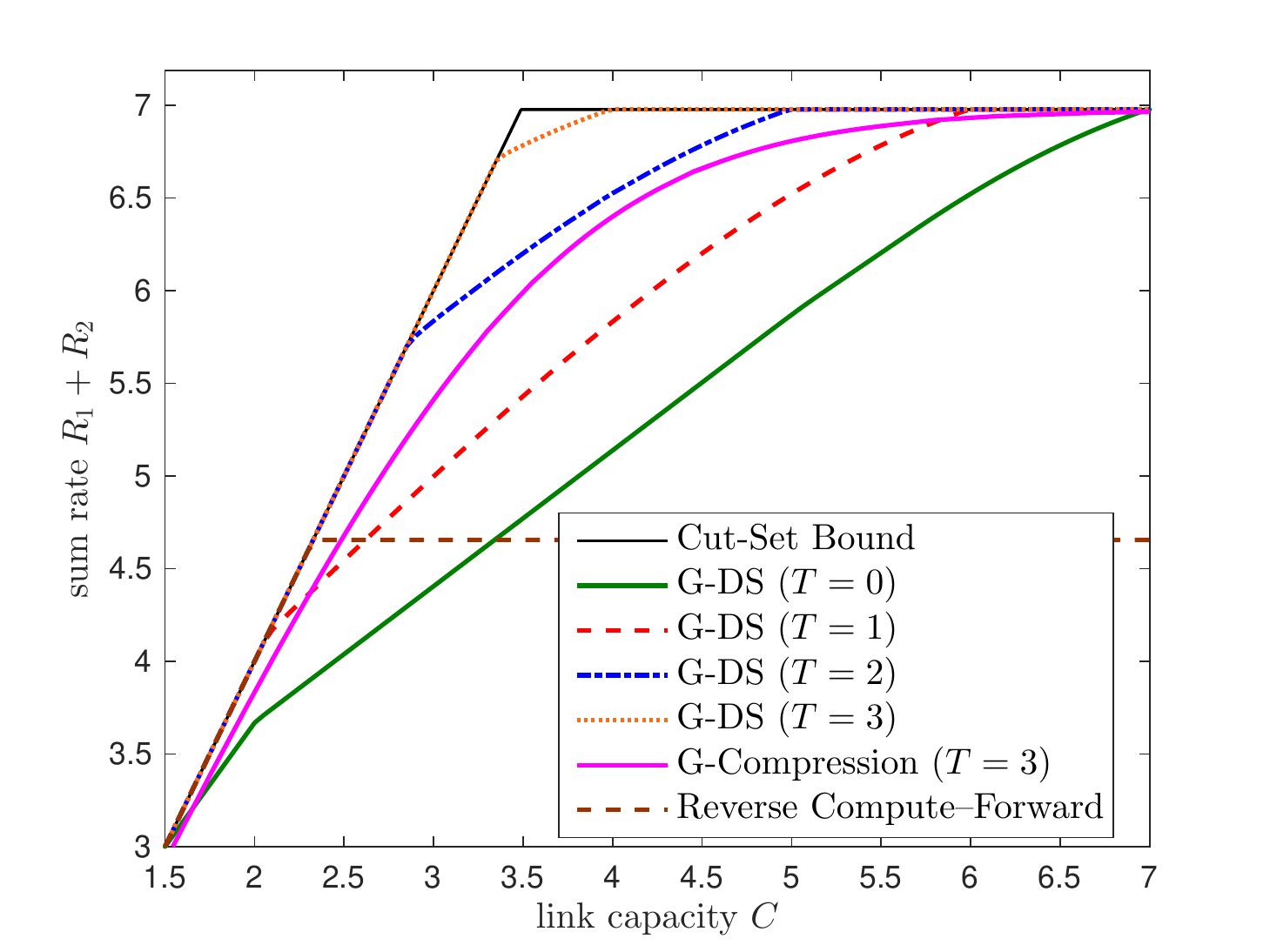}
\end{center}
\vspace{-0.15in}
\caption{Achieved sum-rates of the G-DS scheme, the G-Compression scheme, and the reverse compute--forward scheme with power control under the symmetric memoryless Gaussian model. Here $P=100$ and \usebox{\chmatsym}.}
\label{fig:T0123}
\vspace{-0.1in}
\end{figure}

Finally, we consider BS cooperation, i.e., the case where $T>0$.\footnote{We note that the reverse compute--forward has not been extended for the scenario with BS cooperation. We only include it here as a reference.} Figure \ref{fig:T0123} plots the achieved sum rates for the case of $(P,g_{12},g_{21})=(100,0.5,-0.5)$. It turns out that for the symmetric case, only the G-DS scheme can benefit from the cooperation links. In particular, as the link capacity $T$ increases to two, the G-DS scheme already outperforms the G-Compression scheme for all values of $C$. Recall that the G-DS scheme~I achieves the sum rate $\min\{C+T,2C,R_{\sf sum}^\star\}$. Since the cut-set bound is $\min\{2C,R_{\sf sum}^\star\}$, we see that increasing $T$ is beneficial when $R_1+R_2<C+T$ is the dominating constraint. By contrast, for the symmetric case the G-Compression scheme cannot benefit from the cooperation links because the dominating rate constraints do not involve $C_{12}$ and $C_{21}$: 
\begin{IEEEeqnarray}{rCl} 
R_1+R_2 &<& I(U_1;Y_1)+I(U_2;Y_2)-I(U_1;U_2) \IEEEnonumber \\
&& + \min\left\{\begin{array}{l}
0, \\
C_1+C_2-I(U_1,U_2;X_0,X_1,X_2)-I(X_1;X_2|X_0)   
\end{array}\right\}, \label{eq:dm_cst}
\end{IEEEeqnarray}
which can be rewritten as 
\begin{IEEEeqnarray*}{rCl}
R_1+R_2 &<& I(U_1;Y_1)+I(U_2;Y_2)-I(U_1;U_2), \\
R_1+R_2 &<& C_1+C_2 - I(U_1;X_0,X_1,X_2|Y_1) - I(U_2;U_1,X_0,X_1,X_2|Y_2) -I(X_1;X_2|X_0).
\end{IEEEeqnarray*}

\newsavebox{\asymchmat}
\savebox{\asymchmat}{$\left[\begin{smallmatrix}g_{11}  & g_{12} \\ g_{21} & g_{22}\end{smallmatrix}\right]=\left[\begin{smallmatrix}1&0.25\\1&-0.25\end{smallmatrix}\right]$}

\begin{figure}[t!]
\vspace{-0.1in}
\begin{center}
\includegraphics[scale=0.8]{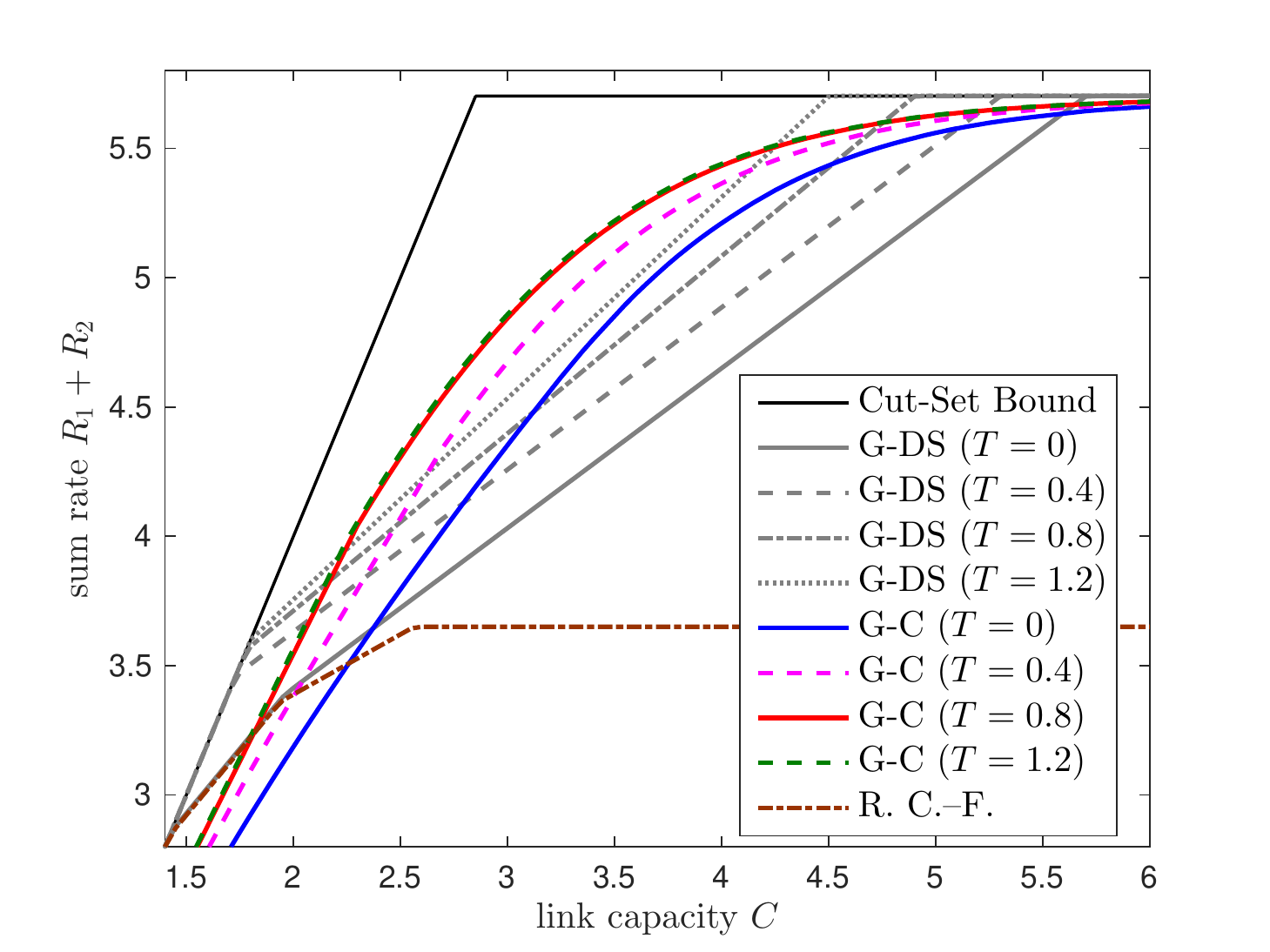}
\end{center}
\vspace{-0.15in}
\caption{Achieved sum-rates of the G-DS scheme, the G-Compression (G-C) scheme, and the reverse compute--forward scheme with power control (R. C.-F.) under the memoryless Gaussian model. Here $P=100$ and \usebox{\asymchmat}.}
\label{fig:asym}
\vspace{-0.1in}
\end{figure}

If the channel gain matrix is asymmetric, the G-Compression scheme can benefit from the cooperation links, but the gain eventually saturates as $T$ increases, again due to the dominating constraint~\eqref{eq:dm_cst}. Figure~\ref{fig:asym} plots the achieved sum rates for the case of $P=100$ and \usebox{\asymchmat}. As can been seen, as $T$ increases from $0.8$ to $1.2$, there is little improvement for the G-Compression scheme. By contrast, the G-DS scheme keep benefiting from the cooperation links before coinciding with the cut-set bound, especially when the link capacity $C$ is large.

\appendices

\section{Expected Size of Independently Generated Codebooks} \label{apdx:ind}
The following lemma is a simple extension of \cite[Problem 3.8, p. 73]{ElGamal:11} (see also \cite{Traskov:07}).
\begin{lemma} \label{lma:ind}
Let $(U,V,W)\sim p_{U,V,W}$. Let $W^n$ be generated according to $\prod_{i=1}^np_{W}(w_i)$. Consider two independently generated codebooks $\mathcal{C}_1=\{U^n(1),\cdots,U^n(2^{nR_1})\}$ and $\mathcal{C}_2=\{V^n(1),\cdots,V^n(2^{nR_2})\}$. The codewords of $\mathcal{C}_1$ are generated independently each according to $\prod_{i=1}^np_{U}(u_i)$. The codewords of $\mathcal{C}_2$ are generated independently each according to $\prod_{i=1}^np_{V}(v_i)$. Define the set 
\begin{IEEEeqnarray*}{rCl}
\mathcal{C} &=& \{(u^n,v^n)\in\mathcal{C}_1\times\mathcal{C}_2:(u^n,v^n,W^n)\in\mathcal{T}_\epsilon^{(n)}(U,V,W)\}.
\end{IEEEeqnarray*}
Then, there exists $\delta(\epsilon)>0$ that tends to zero as $\epsilon\to 0$ such that 
\begin{IEEEeqnarray*}{rCl}
\mathbb{E}[|\mathcal{C}|] &\le& 2^{n(R_1+R_2-I(U;V)-I(U,V;W)+\delta(\epsilon))}. 
\end{IEEEeqnarray*}

\end{lemma}
\begin{IEEEproof}
\begin{IEEEeqnarray*}{rCl}
\mathbb{E}[|\mathcal{C}|] &=& \sum_{m=1}^{2^{nR_1}}\sum_{\ell=1}^{2^{nR_2}}\mathbb{P}((U^n(m),V^n(\ell),W^n)\in\mathcal{T}_\epsilon^{(n)}) \\
&=& 2^{n(R_1+R_2)}\mathbb{P}((U^n(1),V^n(1),W^n)\in\mathcal{T}_\epsilon^{(n)}) \\
&=& 2^{n(R_1+R_2)}\sum_{w^n\in\mathcal{T}_\epsilon^{(n)}}p_{W^n}(w^n)\sum_{(u^n,v^n)\in\mathcal{T}_\epsilon^{(n)}(U,V|w^n)}p_{U^n}(u^n)p_{V^n}(v^n) \\
&\le& 2^{n(R_1+R_2)}\sum_{w^n\in\mathcal{T}_\epsilon^{(n)}}p_{W^n}(w^n)\sum_{(u^n,v^n)\in\mathcal{T}_\epsilon^{(n)}(U,V|w^n)}2^{-n(H(U)-\delta(\epsilon))}2^{-n(H(V)-\delta(\epsilon))} \\
&=& 2^{n(R_1+R_2)} \sum_{w^n\in\mathcal{T}_\epsilon^{(n)}}p_{W^n}(w^n) |T_\epsilon^{(n)}(U,V|w^n)| 2^{-n(H(U)+H(V)-2\delta(\epsilon))} \\
&\le& 2^{n(R_1+R_2)} 2^{n(H(U,V|W)+\delta(\epsilon))} 2^{-n(H(U)+H(V)-2\delta(\epsilon))} \\
&=& 2^{n(R_1+R_2-I(U;V)-I(U,V;W)+3\delta(\epsilon))}.
\end{IEEEeqnarray*}
\end{IEEEproof}

\section{Multivariate Covering Lemma with Non-Cartesian Product Sets} \label{apdx:smcl}
\begin{lemma} \label{lma:smcl}
Let $(U_0,U_1,U_2,V_0,V_1,V_2)\sim p_{U_0,U_1,U_2,V_0,V_1,V_2}$. For $j\in\{0,1,2\}$, randomly and independently generate sequences $U_j^n(k_j)$, $k_j\in[2^{nR_{{\sf u}j}}]$ , each according to $\prod_{i=1}^np_{U_j}(u_{ji})$. For $j\in\{0,1,2\}$, randomly and independently generate sequences $V_j^n(\ell_j)$, $\ell_j\in[2^{nR_{{\sf v}j}}]$ , each according to $\prod_{i=1}^np_{V_j}(v_{ji})$. Randomly and independently assign an index $m_1(k_0,k_1,k_2)$ to each index tuple $(k_0,k_1,k_2)\in[2^{nR_{{\sf u}0}}]\times[2^{nR_{{\sf u}1}}]\times[2^{nR_{{\sf u}2}}]$ according to a uniform pmf over $[2^{nR_1}]$. Randomly and independently assign an index $m_2(\ell_0,\ell_1,\ell_2)$ to each index tuple $(\ell_0,\ell_1,\ell_2)\in[2^{nR_{{\sf v}0}}]\times[2^{nR_{{\sf v}1}}]\times[2^{nR_{{\sf v}2}}]$ according to a uniform pmf over $[2^{nR_2}]$. 
Denote 
\begin{IEEEeqnarray*}{rCl}
\mathcal{E}(m_1,m_2) &=& \{(U_0^n(k_0),U_1^n(k_1),U_2^n(k_2),V_0^n(\ell_0),V_1^n(\ell_1),V_2^n(\ell_2))\notin\mathcal{T}_\epsilon^{(n)} \\
&& \text{ for all } (k_0,k_1,k_2)\in\mathcal{B}_1(m_1), (\ell_0,\ell_1,\ell_2)\in\mathcal{B}_2(m_2)\}.
\end{IEEEeqnarray*}
For each $(m_1,m_2)\in[2^{nR_1}]\times[2^{nR_2}]$, there exists $\delta(\epsilon)$ that tends to zero as $\epsilon\to 0$ such that 
$\lim_{n\to\infty}\mathbb{P}(\mathcal{E}(m_1,m_2))=0$, if 
\begin{IEEEeqnarray*}{rCl}
\label{eq:smcl:cond}
\sum_{i\in\Omega_{\sf u}} R_{{\sf u}i} + \sum_{j\in\Omega_{\sf v}} R_{{\sf v}j} 
&>& \mathbbold{1}\{\Omega_{\sf u}=\{0,1,2\}\}R_1 + \mathbbold{1}\{\Omega_{\sf v}=\{0,1,2\}\}R_2 + \Gamma(U(\Omega_{\sf u}),V(\Omega_{\sf v})), 
\end{IEEEeqnarray*}
for all $\Omega_{\sf u}, \Omega_{\sf v}\subseteq\{0,1,2\}$ such that $|\Omega_{\sf u}|+|\Omega_{\sf v}|\ge 2$.
\end{lemma}

\begin{IEEEproof}
The proof follows similar steps as the proof of the multivariate covering lemma. The only difference is that now the set of index tuples is not the usual Cartesian product. By symmetry, it suffices to investigate the case $(m_1,m_2)=(1,1)$. For notational convenience, hereafter we denote $\mathcal{B}_j(1)=\mathcal{B}_j$, $j\in\{1,2\}$.

Let 
\begin{IEEEeqnarray*}{rCl}
\mathcal{A} &=& \{(k_0,k_1,k_2,\ell_0,\ell_1,\ell_2):(U_0^n(k_0),U_1^n(k_1),U_2^n(k_2),V_0^n(\ell_0),V_1^n(\ell_1),V_2^n(\ell_2))\in\mathcal{T}_\epsilon^{(n)}, \\
&& (k_0,k_1,k_2)\in\mathcal{B}_1,(\ell_0,\ell_1,\ell_2)\in\mathcal{B}_2\}.
\end{IEEEeqnarray*}
Then, we have 
\begin{IEEEeqnarray*}{rCl}
\mathbb{P}(\mathcal{E}(1,1)) &=& \mathbb{P}(|\mathcal{A}|=0) \\
&\le& \mathbb{P}\left((|\mathcal{A}|-\mathbb{E}[|\mathcal{A}|])^2\ge \mathbb{E}[|\mathcal{A}|]^2\right) \\
\label{eq:mcl_base1}
&\overset{(a)}{\le}& \frac{\text{Var}(|\mathcal{A}|)}{\mathbb{E}[|\mathcal{A}|]^2}
\end{IEEEeqnarray*}
where $(a)$ follows from Chebyshev's inequality. For convenience, denote 
\begin{IEEEeqnarray*}{rCl}
\phi(k_0,k_1,k_2,\ell_0,\ell_1,\ell_2) &=& \mathbbold{1}\{(U_0^n(k_0),U_1^n(k_1),U_2^n(k_2),V_0^n(\ell_0),V_1^n(\ell_1),V_2^n(\ell_2))\in\mathcal{T}_\epsilon^{(n)}\}.
\end{IEEEeqnarray*}
Then, the set size $|\mathcal{A}|$ conditioned on the random bin assignments $\mathcal{B}_1$ and $\mathcal{B}_2$ can be expressed as 
\begin{IEEEeqnarray*}{rCl}
\mathbb{E}[|\mathcal{A}||\mathcal{B}_1,\mathcal{B}_2] &=& \sum_{(k_1,k_2)\in\mathcal{B}_1}\sum_{(\ell_1,\ell_2)\in\mathcal{B}_2}\phi(k_0,k_1,k_2,\ell_0,\ell_1,\ell_2).
\end{IEEEeqnarray*} 

For $a_0,a_1,a_2,b_0,b_1,b_2\in\{1,2\}$, let 
\begin{IEEEeqnarray*}{ll}
& p(a_0,a_1,a_2,b_0,b_1,b_2) \\
&= \mathbb{E}[\phi(1,1,1,1,1,1)\phi(a_0,a_1,a_2,b_0,b_1,b_2)], \\
& Q(a_0,a_1,a_2,b_0,b_1,b_2) \\
&= |\{(k_0,k_1,k_2,\ell_0,\ell_1,\ell_2,k'_0,k'_1,k'_2,\ell'_0,\ell'_1,\ell'_2):  \\
& \hspace{0.6cm}(k_0,k_1,k_2)\in\mathcal{B}_1,(\ell_0,\ell_1,\ell_2)\in\mathcal{B}_2,(k'_0,k'_1,k'_2)\in\mathcal{B}_1,(\ell'_0,\ell'_1,\ell'_2)\in\mathcal{B}_2,  \\
& \hspace{0.6cm} \mathcal{F}_0^{(a_0)},\mathcal{F}_1^{(a_1)},\mathcal{F}_2^{(a_2)},\mathcal{G}_0^{(b_0)},\mathcal{G}_1^{(b_1)},\mathcal{G}_2^{(b_2)}\}|,
\end{IEEEeqnarray*}
where $\mathcal{F}_j^{(1)} = \left(\mathcal{F}_j^{(2)}\right)^c = \{k_j=k'_j\}$ and $\mathcal{G}_j^{(1)} = \left(\mathcal{G}_j^{(2)}\right)^c = \{\ell_j=\ell'_j\}$, for $j\in\{0,1,2\}$. 
Then, we have
\begin{IEEEeqnarray*}{ll}
& \mathbb{E}[|\mathcal{A}||\mathcal{B}_1,\mathcal{B}_2] \\
&= \sum_{(k_0,k_1,k_2)\in\mathcal{B}_1}\sum_{(\ell_0,\ell_1,\ell_2)\in\mathcal{B}_2}\mathbb{E}[\phi(k_0,k_1,k_2,\ell_0,\ell_1,\ell_2)] \\
&= Q(1,1,1,1,1,1)p(1,1,1,1,1,1), \\
& \mathbb{E}[|\mathcal{A}|^2|\mathcal{B}_1,\mathcal{B}_2] \\
&= \sum_{(k_0,k_1,k_2)\in\mathcal{B}_1}\sum_{(\ell_0,\ell_1,\ell_2)\in\mathcal{B}_2}\sum_{(k'_0,k'_1,k'_2)\in\mathcal{B}_1}\sum_{(\ell'_0,\ell'_1,\ell'_2)\in\mathcal{B}_2}\mathbb{E}[\phi(k_0,k_1,k_2,\ell_0,\ell_1,\ell_2)\phi(k'_0,k'_1,k'_2,\ell'_0,\ell'_1,\ell'_2)] \IEEEeqnarraynumspace\\
&= \sum_{a_0,a_1,a_2,b_0,b_1,b_2}Q(a_0,a_1,a_2,b_0,b_1,b_2)p(a_0,a_1,a_2,b_0,b_1,b_2).
\end{IEEEeqnarray*}
Hence 
\begin{IEEEeqnarray*}{rCl}
\frac{\text{Var}(|\mathcal{A}|)}{\mathbb{E}[|\mathcal{A}|]^2} &=& 
\frac{\mathbb{E}[\mathbb{E}[|\mathcal{A}|^2|\mathcal{B}_1,\mathcal{B}_2]]-(\mathbb{E}[\mathbb{E}[|\mathcal{A}||\mathcal{B}_1,\mathcal{B}_2]])^2}{(\mathbb{E}[\mathbb{E}[|\mathcal{A}||\mathcal{B}_1,\mathcal{B}_2]])^2}  \\
&=& \frac{\displaystyle\sum_{(a_0,a_1,a_2,b_0,b_1,b_2)\neq(2,2,2,2,2,2)}\mathbb{E}[Q(a_0,a_1,a_2,b_0,b_1,b_2)]p(a_0,a_1,a_2,b_0,b_1,b_2)}{(\mathbb{E}[Q(1,1,1,1,1,1)]p(1,1,1,1,1,1))^2} \label{eq:mcl_base2}
\end{IEEEeqnarray*}

Denote $I=\Gamma(U_0,U_1,U_2,V_0,V_1,V_2)$. By the joint typicality lemma~\cite[p. 29]{ElGamal:11}, it holds that 
\begin{IEEEeqnarray*}{rCl}
\label{eq:mcl_cond1}
p(1,1,1,1,1,1) &\ge& 2^{-n(I+\delta(\epsilon))}, \\
\label{eq:mcl_cond2}
p(a_0,a_1,a_2,b_0,b_1,b_2) &\le& 2^{-n(I+\sum_{i\in\Omega_{\sf u}^c}H(U_i)+\sum_{j\in\Omega_{\sf v}^c}H(V_j)-H(U(\Omega_{\sf u}^c),V(\Omega_{\sf v}^c)|U(\Omega_{\sf u}),V(\Omega_{\sf v}))-\delta(\epsilon))}, 
\end{IEEEeqnarray*}
where $\Omega_{\sf u} = \bigcup_{j=0}^2\kappa_j(a_j)$, $\Omega_{\sf v} = \bigcup_{j=0}^2\kappa_j(b_j)$, and 
\begin{IEEEeqnarray*}{rCl}
\kappa_j(x) &=& \begin{cases}
\{j\} & \text{ if }  x=1, \\
\emptyset & \text{ otherwise.} 
\end{cases}
\end{IEEEeqnarray*}
Also, for all $a_0,a_1,a_2,b_0,b_1,b_2\in\{1,2\}$, we have 
\begin{IEEEeqnarray*}{rCl}
\label{eq:mcl_cond3}
\mathbb{E}[Q(a_0,a_1,a_2,b_0,b_1,b_2)] = 2^{n\left(\sum_{i=0}^2a_iR_{{\sf u}i}+\sum_{j=0}^2b_jR_{{\sf v}j}-(1+\mathbbold{1}\{\bigcup_{i=0}^2\{a_i=2\})R_1-(1+\mathbbold{1}\{\bigcup_{j=0}^2\{b_j=2\})R_2\right)}.\IEEEeqnarraynumspace
\end{IEEEeqnarray*}

Finally, \eqref{eq:mcl_base2} and thus \eqref{eq:mcl_base1} can be further upper bounded using \eqref{eq:mcl_cond1},\eqref{eq:mcl_cond2},\eqref{eq:mcl_cond3}. It can be checked that the corresponding upper bound tends to zero as $n\to\infty$ if the condition \eqref{eq:smcl:cond} holds, which establishes the lemma.
\end{IEEEproof}

\section{Proof of Proposition \ref{prop:DDF}} \label{appendix:DDF}
We first establish the achievability. To achieve \eqref{eq:DDF_refined}, we set $p_{\tilde{U}^N,U^L,\tilde{X},(X^N,\{W_{kj}\})}=p_{\tilde{U}^N,\tilde{X},\{W_{kj}\}}p_{U^L,X^N}$, where  
\begin{enumerate}
\item $\tilde{U}_k=(W_k,(W_{kj}:j\neq k))$; 
\item $p_{W^N,\{W_{kj}\}}=\prod_{k=1}^Np_{W_k}\prod_{(j,k):j\neq k}p_{W_{kj}}$;
\item $W_k\sim$ Uniform($[2^{C_k}]$); and 
\item $W_{kj}\sim$ Uniform($[2^{C_{kj}}]$).
\end{enumerate}

Next, we show that the rate expression in \eqref{eq:DDF} can be upper bounded by the rate expression in \eqref{eq:DDF_refined} and thus establish the converse. Indeed, we have 
\begin{IEEEeqnarray*}{ll}
& I(\tilde{X},\breve{X}(\mathcal{S});\tilde{U}(\mathcal{S}^c),U(\mathcal{D})|\breve{X}(\mathcal{S}^c)) 
- \sum_{k\in\mathcal{S}^c}\left[ I(\tilde{U}_k;\tilde{U}(\mathcal{S}_k^c),\tilde{X},\breve{X}^N|\breve{X}_k,\breve{Y}_k)+I(\breve{X}_k;\breve{X}(\mathcal{S}_k^c))\right]  \\
& \hspace{0.3cm} -\sum_{\ell\in\mathcal{D}} I(U_\ell;U(\mathcal{D}_\ell),\tilde{U}(\mathcal{S}^c),\tilde{X},\breve{X}^N|Y_\ell) \\
&\overset{(a)}{=} I(\tilde{X},\breve{X}^N;\tilde{U}(\mathcal{S}^c),U(\mathcal{D})|\breve{X}(\mathcal{S}^c)) - \sum_{k\in\mathcal{S}^c}\left[ I(\tilde{U}_k;\tilde{U}(\mathcal{S}_k^c),\tilde{X},\breve{X}^N) - I(\tilde{U}_k;\breve{Y}_k,\breve{X}_k)\right]  \\
& \hspace{0.3cm} - \sum_{k\in\mathcal{S}^c} I(\breve{X}_k;\breve{X}(\mathcal{S}_k^c)) -\sum_{\ell\in\mathcal{D}} \left[I(U_\ell;U(\mathcal{D}_\ell),\tilde{U}(\mathcal{S}^c),\tilde{X},\breve{X}^N) - I(U_\ell;Y_\ell) \right] \\
&=\sum_{\ell\in\mathcal{D}}I(U_\ell;Y_\ell) + \sum_{k\in\mathcal{S}^c} I(\tilde{U}_k;\breve{Y}_k,\breve{X}_k) + I(\tilde{X},\breve{X}(\mathcal{S});\tilde{U}(\mathcal{S}^c),U(\mathcal{D})|\breve{X}(\mathcal{S}^c))  \\
& \hspace{0.3cm} - \sum_{k\in\mathcal{S}^c} I(\breve{X}_k;\breve{X}(\mathcal{S}_k^c)) - \left[\sum_{k\in\mathcal{S}^c} H(\tilde{U}_k) + \sum_{\ell\in\mathcal{D}} H(U_\ell) - H(\tilde{U}(\mathcal{S}^c),U(\mathcal{D})|\tilde{X},\breve{X}^N) \right] \\
&=\sum_{\ell\in\mathcal{D}}I(U_\ell;Y_\ell) + \sum_{k\in\mathcal{S}^c} I(\tilde{U}_k;\breve{Y}_k|\breve{X}_k)  \\
& \hspace{0.3cm} - \sum_{k\in\mathcal{S}^c} I(\breve{X}_k;\breve{X}(\mathcal{S}_k^c)) - \left[ \sum_{k\in\mathcal{S}^c} H(\tilde{U}_k|\breve{X}_k) +\sum_{\ell\in\mathcal{D}} H(U_\ell) - H(\tilde{U}(\mathcal{S}^c),U(\mathcal{D})|\breve{X}(\mathcal{S}^c)) \right] \\
&=\sum_{\ell\in\mathcal{D}}I(U_\ell;Y_\ell) + \sum_{k\in\mathcal{S}^c} I(\tilde{U}_k;\breve{Y}_k|\breve{X}_k) - \left[ \sum_{k\in\mathcal{S}^c} H(\tilde{U}_k|\breve{X}_k) - H(\tilde{U}(\mathcal{S}^c)|X(\mathcal{S}^c),U(\mathcal{D})) \right]  \\
& \hspace{0.3cm} - \sum_{k\in\mathcal{S}^c} I(\breve{X}_k;\breve{X}(\mathcal{S}_k^c)) - \left[ \sum_{\ell\in\mathcal{D}} H(U_\ell) - H(U(\mathcal{D})|\breve{X}(\mathcal{S}^c)) \right] \\
&\le \sum_{\ell\in\mathcal{D}}I(U_\ell;Y_\ell) + \sum_{k\in\mathcal{S}^c} H(\breve{Y}_k) - \sum_{k\in\mathcal{S}^c} I(\breve{X}_k;\breve{X}(\mathcal{S}_k^c)) - \left[ \sum_{\ell\in\mathcal{D}} H(U_\ell) - H(U(\mathcal{D})|\breve{X}(\mathcal{S}^c)) \right] \\
&\le \sum_{\ell\in\mathcal{D}}I(U_\ell;Y_\ell) + \sum_{k\in\mathcal{S}^c} C_k + \sum_{j\in\mathcal{S}}\sum_{k\in\mathcal{S}^c} C_{kj} - \sum_{k\in\mathcal{S}^c} I(\breve{X}_k;\breve{X}(\mathcal{S}_k^c)) - \sum_{\ell\in\mathcal{D}} I(U_\ell;U(\mathcal{D}_\ell),\breve{X}(\mathcal{S}^c)) \\
&\le \sum_{\ell\in\mathcal{D}}I(U_\ell;Y_\ell) + \sum_{k\in\mathcal{S}^c} C_k + \sum_{j\in\mathcal{S}}\sum_{k\in\mathcal{S}^c} C_{kj} - \sum_{k\in\mathcal{S}^c} I(X_k;X(\mathcal{S}_k^c)) - \sum_{\ell\in\mathcal{D}} I(U_\ell;U(\mathcal{D}_\ell),X(\mathcal{S}^c)), \IEEEeqnarraynumspace
\end{IEEEeqnarray*}
where $(a)$ follows since $\breve{Y}_k$ is a function of $(\tilde{X},X^N)$ and since $(\tilde{X},\tilde{U}^N,U^L)\markov X^N \markov Y^L$ form a Markov chain. Finally, we note that 
\begin{IEEEeqnarray*}{rCl}
\sum_{k\in\mathcal{S}^c} I(X_k;X(\mathcal{S}_k^c)) + \sum_{\ell\in\mathcal{D}} I(U_\ell;U(\mathcal{D}_\ell),X(\mathcal{S}^c)) &=& \Gamma(X(\mathcal{S}^c),U(\mathcal{D})). 
\end{IEEEeqnarray*}

\section{Proof of Theorem \ref{thm:gap}} \label{appendix:gap}
First, we state the cut-set bound for the capacity region of the memoryless Gaussian C-RAN model. The proof follows by applying the standard cut-set argument (see \cite[Theorem 15.10.1]{Cover:06}) to the considered model and then specializing it to the memoryless Gaussian case. 


\begin{proposition}
If a rate tuple $(R_1,\cdots,R_L)$ is achievable for the downlink $N$-BS $L$-user C-RAN with BS cooperation, then it must satisfy the inequality
\begin{IEEEeqnarray*}{rCl}
\sum_{\ell\in\mathcal{D}} R_\ell &\le& \sum_{k\in\mathcal{S}^c} C_k + \sum_{j\in\mathcal{S}}\sum_{k\in\mathcal{S}^c} C_{kj} + \frac{1}{2}\log\det\left(\mathsf{I}+\mathsf{G}(\mathcal{D},\mathcal{S})\mathsf{K}(\mathcal{S}|\mathcal{S}^c)\mathsf{G}^T(\mathcal{D},\mathcal{S})\right),
\end{IEEEeqnarray*}
for all $\mathcal{S}\subseteq[N]$ and all nonempty subsets $\mathcal{D}\subseteq[L]$ for some covariance matrix $\mathsf{K}\succeq 0$ with $\mathsf{K}_{jj}\le P$. Here $\mathsf{K}(\mathcal{S}|\mathcal{S}^c)$ is the conditional covariance matrix of $X(\mathcal{S})$ given $X(\mathcal{S}^c)$ for $X^N\sim \mathcal{N}(0,\mathsf{K})$ and $\mathsf{G}(\mathcal{S},\mathcal{D})$ is defined such that 
\begin{IEEEeqnarray*}{rCl}
\begin{bmatrix}Y(\mathcal{D}) \\ Y(\mathcal{D}^c) \end{bmatrix} &=& \begin{bmatrix}\mathsf{G}(\mathcal{D},\mathcal{S}) & \mathsf{G}(\mathcal{D},\mathcal{S}^c) \\ \mathsf{G}(\mathcal{D}^c,\mathcal{S}) & \mathsf{G}(\mathcal{D}^c,\mathcal{S}^c) \end{bmatrix}\begin{bmatrix}X(\mathcal{S}) \\ X(\mathcal{S}^c) \end{bmatrix} + \begin{bmatrix}Z(\mathcal{D}) \\ Z(\mathcal{D}^c) \end{bmatrix}.
\end{IEEEeqnarray*}
\end{proposition}

Now we are ready to prove Theorem \ref{thm:gap}. First, note that \eqref{eq:DDF_refined} can also be expressed as 
\begin{IEEEeqnarray*}{rCl}
\sum_{\ell\in\mathcal{D}} R_\ell &<& \sum_{k\in\mathcal{S}^c} C_k + \sum_{j\in\mathcal{S}}\sum_{k\in\mathcal{S}^c} C_{kj} + I(X(\mathcal{S});U(\mathcal{D})|X(\mathcal{S}^c)) \\ 
&&  - \sum_{k\in\mathcal{S}^c} I(X_k;X(\mathcal{S}_k^c)) 
 -\sum_{\ell\in\mathcal{D}} I(U_\ell;U(\mathcal{D}_\ell),X^N|Y_\ell),
\end{IEEEeqnarray*}
Then, we set $X_k$ to be i.i.d. $\mathcal{N}(0,P)$ for all $k\in[N]$ and 
\begin{IEEEeqnarray*}{rCl}
U_\ell &=& \sum_{k=1}^N g_{\ell k} X_k + \hat{Z}_\ell,
\end{IEEEeqnarray*}
where $\hat{Z}_\ell\sim \mathcal{N}(0,1)$ are mutually independent and independent of $(X^N,Y^L)$. Then, we have 
\begin{IEEEeqnarray*}{rCl}
\sum_{\ell\in\mathcal{D}} R_\ell &<& \sum_{k\in\mathcal{S}^c} C_k + \sum_{j\in\mathcal{S}}\sum_{k\in\mathcal{S}^c} C_{kj} + \frac{1}{2}\log\det\left(\mathsf{I}+P\mathsf{G}(\mathcal{D},\mathcal{S})\mathsf{G}^T(\mathcal{D},\mathcal{S})\right) \\
&& -\sum_{\ell\in\mathcal{D}} \frac{1}{2}\log\left(1+\frac{\sum_{k\in\mathcal{S}} g_{\ell k}^2P}{1+\sum_{k\in\mathcal{S}} g_{\ell k}^2P}\right), 
\end{IEEEeqnarray*}
which can be further relaxed as 
\begin{IEEEeqnarray*}{rCl}
\label{eq:inner_relax}
\sum_{\ell\in\mathcal{D}} R_\ell &<& \sum_{k\in\mathcal{S}^c} C_k + \sum_{j\in\mathcal{S}}\sum_{k\in\mathcal{S}^c} C_{kj} + \frac{1}{2}\log\det\left(\mathsf{I}+P\mathsf{G}(\mathcal{D},\mathcal{S})\mathsf{G}^T(\mathcal{D},\mathcal{S})\right)
-\frac{|\mathcal{D}|}{2}. 
\end{IEEEeqnarray*}

On the other hand, the cut-set bound for the Gaussian case is given by  
\begin{IEEEeqnarray*}{rCl}
\sum_{\ell\in\mathcal{D}} R_\ell &\le& \sum_{k\in\mathcal{S}^c} C_k + \sum_{j\in\mathcal{S}}\sum_{k\in\mathcal{S}^c} C_{kj} + \frac{1}{2}\log\det\left(\mathsf{I}+\mathsf{G}(\mathcal{D},\mathcal{S})\mathsf{K}(\mathcal{S}|\mathcal{S}^c)\mathsf{G}^T(\mathcal{D},\mathcal{S})\right), \\
&\overset{(a)}{=}& \sum_{k\in\mathcal{S}^c} C_k + \sum_{j\in\mathcal{S}}\sum_{k\in\mathcal{S}^c} C_{kj} + \frac{1}{2}\log\det\left(\mathsf{I}+\mathsf{G}^T(\mathcal{D},\mathcal{S})\mathsf{G}(\mathcal{D},\mathcal{S})\mathsf{K}(\mathcal{S}|\mathcal{S}^c)\right),
\end{IEEEeqnarray*}
where $(a)$ follows from Sylvester's determinant identity. The term $\det\left(\mathsf{I}+\mathsf{G}^T(\mathcal{D},\mathcal{S})\mathsf{G}(\mathcal{D},\mathcal{S})\mathsf{K}(\mathcal{S}|\mathcal{S}^c)\right)$ can be upper bounded in two different ways. Note that the symmetric matrices $\mathsf{G}^T(\mathcal{D},\mathcal{S})\mathsf{G}(\mathcal{D},\mathcal{S})$ and $\mathsf{K}(\mathcal{S}|\mathcal{S}^c)$ are positive semi-definite. When $\mathcal{S}$ is an empty set, the inner bound matches the cut-set bound. In the following, we consider the case $|\mathcal{S}|\ge 1$.

First, we have 
\begin{IEEEeqnarray*}{rCl}
\det\left(\mathsf{I}+\mathsf{G}^T(\mathcal{D},\mathcal{S})\mathsf{G}(\mathcal{D},\mathcal{S})\mathsf{K}(\mathcal{S}|\mathcal{S}^c)\right) &\le& \det\left(\mathsf{I}+P\mathsf{G}^T(\mathcal{D},\mathcal{S})\mathsf{G}(\mathcal{D},\mathcal{S})\right) \cdot \det\left(\mathsf{I}+\frac{1}{P}\mathsf{K}(\mathcal{S}|\mathcal{S}^c)\right) \\
&\overset{(a)}{\le}& \det\left(\mathsf{I}+P\mathsf{G}(\mathcal{D},\mathcal{S})\mathsf{G}^T(\mathcal{D},\mathcal{S})\right) \cdot 2^{|\mathcal{S}|}, 
\end{IEEEeqnarray*}
where $(a)$ follows from Sylvester's determinant identity and Hadamard's inequality. 

Second, denote by $\lambda_j(A)$ the $j$-th largest eigenvalue of the symmetric matrix $A$. For notational convenience, we denote $\mathsf{G}'=\mathsf{G}^T(\mathcal{D},\mathcal{S})\mathsf{G}(\mathcal{D},\mathcal{S})$ and $\mathsf{K}'=\mathsf{K}(\mathcal{S}|\mathcal{S}^c)$.
Note that the matrix $\mathsf{G}'$ has at most $|\mathcal{D}|$ nonzero eigenvalues and $\lambda_1(\mathsf{K}') \le \tr(\mathsf{K}') \le |\mathcal{S}|P$. Thus, we have 
\begin{IEEEeqnarray*}{rCl}
\det\left(\mathsf{I}+\mathsf{G}'\mathsf{K}'\right) &=& \prod_{i=1}^{|\mathcal{S}|}\left(1+\lambda_i(\mathsf{G}'\mathsf{K}')\right) \\
&\overset{(a)}{\le}& \prod_{i=1}^{|\mathcal{S}|}\left(1+\lambda_i(\mathsf{G}')\lambda_1(\mathsf{K}')\right) \\
&\le& \prod_{i=1}^{|\mathcal{S}|}\left(1+\lambda_i(\mathsf{G}')|\mathcal{S}|P\right) \\
&=& \det\left(\mathsf{I}+|\mathcal{S}|P\mathsf{G}(\mathcal{D},\mathcal{S})\mathsf{G}^T(\mathcal{D},\mathcal{S})\right) \\
&\le& \det\left(\mathsf{I}+P\mathsf{G}(\mathcal{D},\mathcal{S})\mathsf{G}^T(\mathcal{D},\mathcal{S})\right) \cdot |\mathcal{S}|^{|\mathcal{D}|}, 
\end{IEEEeqnarray*}
where $(a)$ follows from \cite[7.3.P16]{Horn:13}. 

To summarize, the cut-set bound can be relaxed as 
\begin{IEEEeqnarray*}{rCl}
\label{eq:outer_relax}
\sum_{\ell\in\mathcal{D}} R_\ell &\le& \sum_{k\in\mathcal{S}^c} C_k + \sum_{j\in\mathcal{S}}\sum_{k\in\mathcal{S}^c} C_{kj} + \det\left(\mathsf{I}+P\mathsf{G}(\mathcal{D},\mathcal{S})\mathsf{G}^T(\mathcal{D},\mathcal{S})\right) + \frac{1}{2}\min\{|\mathcal{S}|,|\mathcal{D}|\log|\mathcal{S}|\}. 
\end{IEEEeqnarray*}
Comparing the relaxed inner bound \eqref{eq:inner_relax} and outer bound \eqref{eq:outer_relax}, we conclude that the DDF scheme achieves within $\min\left\{\frac{L+N}{2},\frac{L+L\log N}{2}\right\}$ bits per dimension from the cut-set bound and thus from the capacity region.

\bibliographystyle{IEEEtran}
\bibliography{IEEEabrv,References_Huawei}
\end{document}